\documentclass[12pt]{iopart}
\usepackage{graphicx}
\IfFileExists{url.sty}{\usepackage{url}}
                      {\newcommand{\url}{\texttt}}

\begin{document}

\review[Lunar Laser Ranging]{Lunar Laser Ranging: The Millimeter Challenge}

\author{T. W. Murphy, Jr.}
\address{Center for Astrophysics and Space Sciences, University of California, San Diego, 9500 Gilman Drive, La Jolla, CA 92093-0424, USA}
\ead{tmurphy@physics.ucsd.edu}

\begin{abstract}
Lunar laser ranging has provided many of the best tests of gravitation
since the first Apollo astronauts landed on the Moon. The march to
higher precision continues to this day, now entering the millimeter
regime, and promising continued improvement in scientific results.
This review introduces key aspects of the technique, details the motivations,
observables, and results for a variety of science objectives, summarizes
the current state of the art, highlights new developments in the field,
describes the modeling challenges, and looks to the future of the
enterprise.
\end{abstract}


\maketitle
\tableofcontents

\section{The LLR concept\label{sec:The-LLR-concept}}

Since 1969, lunar laser ranging (LLR) has provided high-precision
measurements of the Earth-Moon distance, contributing to the foundations
of our knowledge in gravitation and planetary physics. While being
the most evident force of nature, gravity is in fact the weakest of
the fundamental forces, and consequently the most poorly tested by
modern experiments. Einstein's
general relativity---currently our best description of gravity---is
fundamentally incompatible with quantum mechanics and is likely to
be replaced by a more complete theory in the future. A modified theory
would, for example, predict small deviations in the solar system that, if seen,
could have profound consequences for understanding the universe as
a whole.

\begin{figure}
\begin{center}
\includegraphics[height=3in,keepaspectratio]{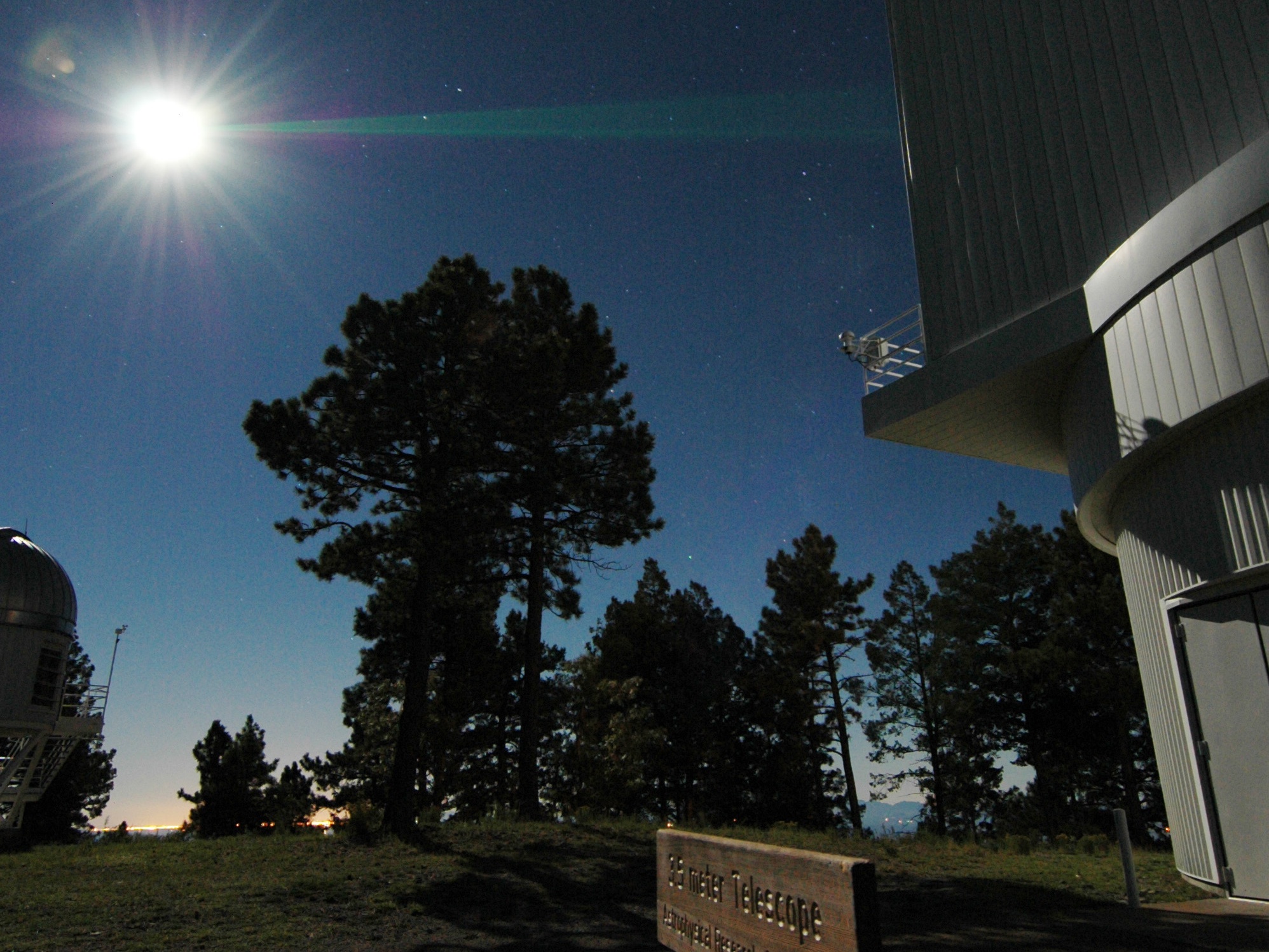}
\end{center}

\caption{{\small Lunar Laser Ranging in action at the Apache Point
Observatory in southern New Mexico.  Photo by Dan Long.\label{fig:APOLLO}}}
\end{figure}

Utilizing reflectors placed on the lunar surface by American astronauts
and Soviet rovers, LLR measures the round-trip travel time of short
pulses of laser light directed to one reflector at a time
(Fig.~\ref{fig:APOLLO}). By mapping
the shape of the lunar orbit, LLR is able to distinguish between competing
theories of gravity. Range precision has improved from a few decimeters
initially to a few millimeters recently, constituting a relative precision
of $10^{-9}$--$10^{-11}$. Leveraging the raw measurement across
the Earth-Sun distance provides another two orders of magnitude for
gauging relativistic effects in the Earth-Moon-Sun system.

As LLR precision has improved over time, the technique has remained at the
cutting edge of tests of gravitational phenomenology and probes of the
lunar interior, and has informed our knowledge of Earth orientation,
precession, and coordinate systems. LLR was last reviewed in this series in
1982 \cite{nord-rop}; this update describes the key science drivers and
findings of LLR, the apparatus and technologies involved, the requisite
modeling techniques, and future prospects on all fronts.  LLR is expected
to continue on its trajectory of improvement, maintaining a leading role in
contributions to science. Other recent reviews by Merkowitz (2010)
\cite{merk-lrr-llr} and by M\" uller et al. (2012) \cite{jog-rev}
complement the present one.  The Merkowitz review, like this one, stresses
gravitational tests of LLR, but with greater emphasis on associated range
signals.  Next-generation reflector and transponder technologies are more
thoroughly covered.  The M\" uller et al. review (for which this author is
a co-author) offers a more complete history of LLR, has statistics on the
LLR data set, and provides greater emphasis on geophysics, selenophysics,
and coordinate systems.

This review is organized as follows: Section~\ref{sec:The-LLR-concept}
provides an overview of the subject; Section\ref{sec:Science-from-LLR}
reviews the science delivered by LLR, with an emphasis on gravitation;
Section~\ref{sec:LLR-capability} describes current LLR capabilities;
Section~\ref{sec:Recent-Surprises} relates recent surprises from LLR,
including the finding of the lost Lunokhod~1 reflector and evidence for
dust accumulation on the reflectors;
Section~\ref{sec:The-Modeling-Challenge} treats the modeling challenges
associated with millimeter-level LLR accuracy; and
Section~\ref{sec:Future-Advances} offers possible future directions for the
practice of LLR.  An Appendix contains a list of acronyms used in the text.
Some single-use ancillary acronyms are only defined in the Appendix in
order to minimize unimportant interruptions.

\subsection{Current Science Results\label{sub:Current-Science-Results}}

A detailed description of the science capabilities of LLR is deferred
until Section~\ref{sec:Science-from-LLR}. For the purposes of introducing
the motivation behind the effort, LLR provides the following leading
tests and measurements:

\begin{itemize}
\item the strong equivalence principle to $\eta\approx3\times10^{-4}$ sensitivity
\cite{jgw-cqg-2012,muller-cqg-2012};
\item time-rate-of-change of the gravitational constant to $\dot{G}/G<10^{-12}$
yr$^{-1}$ \cite{mueller-gdot,jgw-2004,hofmann-2010};
\item geodetic precession within 0.3\% of general relativity prediction
\cite{jog-rev};
\item gravitomagnetism within $\sim0.2$\% of general relativity prediction
\cite{gravmag-prl,soffel};
\item the $1/r^{2}$ law to $\sim2\times 10^{-11}$ times the strength
of gravity at $10^{8}$~m scales \cite{muller-isq,ega-isl};
\item the presence of a liquid core in the Moon having a radius of $\sim350$~km
\cite{liquid-core,khan-core}.
\end{itemize}
LLR also provides checks on preferred frame effects \cite{mueller-a1,nordtvedt-a2},
and Newton's third law \cite{newtonsthird}. LLR may additionally
open a window into the possible existence of extra dimensions via
cosmological dilution of gravity \cite{lue,dvalimoon}. Besides the
strong equivalence principle, LLR tests the weak equivalence principle
at the level of $\Delta a/a<1.3\times10^{-13}$ \cite{jgw-ep}. Laboratory
tests of the weak equivalence principle reach similar levels \cite{npl},
but result in more incisive tests by having the freedom to choose
more optimal mass pairs than the iron-silicate pairing dictated by
the Earth-Moon system. Finally, LLR is used to define coordinate systems,
probe the lunar interior, and study geodynamics \cite{dickey}. Order-of-magnitude
advances in each of these domains is possible as the LLR technique
improves from the centimeter to the millimeter regime.

\subsection{A Quantitative Introduction\label{sub:A-Quantitative-Introduction}}

By way of introduction, this section presents representative numbers
meant to portray key aspects of the LLR technique. Some items will
enter in subsequent discussion, while others appear only here for
the sake of overall familiarity.

While the semi-major axis of the lunar orbit is 384,402~km, the time-averaged
distance between Earth and Moon centers is 385,000.6~km, corresponding
to 2.56~s of round-trip light travel time. The range varies from
356,500--406,700~km, chiefly due to a 21,000~km amplitude oscillation
(27.55~d period) associated with the elliptical orbit of the Moon
($e=0.055$). Other leading oscillations appear at 3700~km (31.8~d)
and 2955~km (14.76~d) due to solar perturbations. The range rate
between Earth and Moon centers may be as large as 75~m~s$^{-1}$,
while Earth rotation is the dominant range-rate effect, measuring
465~m~s$^{-1}$ at the equator.

\begin{figure}
\begin{center}
\includegraphics[width=5in,keepaspectratio]{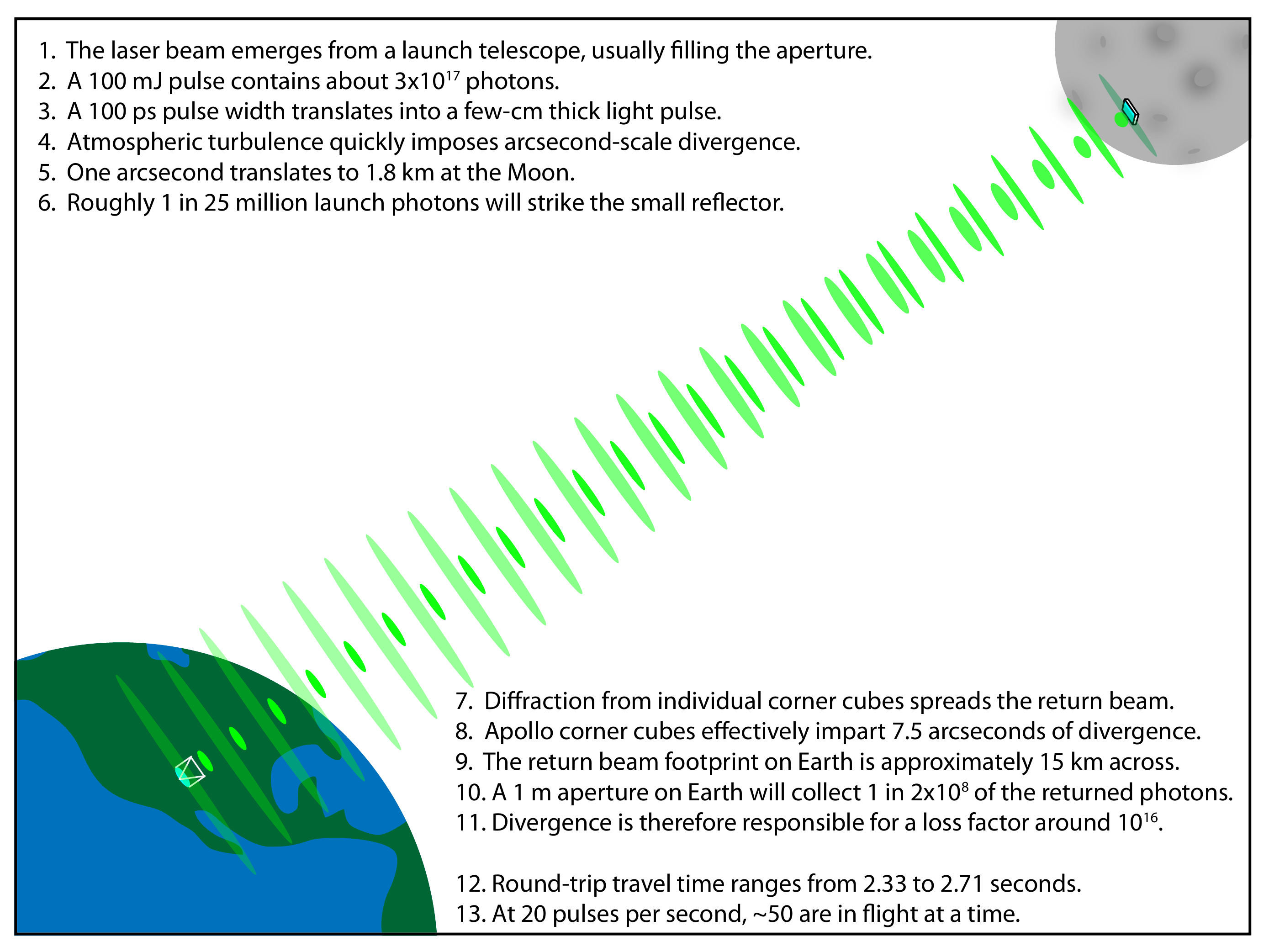}
\end{center}

\caption{{\small Cartoon schematic of LLR technique and divergence-related
challenges. \label{fig:basics}}}
\end{figure}

The basic arrangement for performing lunar laser ranging is shown in
Fig.~\ref{fig:basics}. 
Illuminating the reflectors sufficiently is a principal challenge
in LLR. Even a one-arcsecond (5~$\mu$rad) beam---limited by atmospheric
turbulence---spreads to 1.9~km at the lunar surface. This translates
into a one-in-25-million chance of a photon launched from Earth finding
the Apollo~11 reflector, for instance (discussed in Section~\ref{sub:Reflectors-and-Divergence}).
The return journey is even more difficult, owing to diffractive spread
from the corner cube prisms, compounded by velocity aberration. A
1~m circular aperture on Earth can expect to receive one photon out
of every 250 million emerging from the Apollo reflector. The tangential
relative motion of the Earth station with respect to the Moon introduces
a 4--6~$\mu$rad velocity aberration, translating to a $\sim2$~km
offset of the return pattern on the Earth's surface and a further
reduction of the Apollo reflector signal by a factor of 0.6--0.8. 

A useful conversion to memorize is that 1~mm of separation translates
into 6.67~ps of round-trip travel time, or inversely 1~ns of round-trip
time maps to 0.15~m of one-way distance.

\subsection{Reflectors and Divergence-Imposed Requirements\label{sub:Reflectors-and-Divergence}}

We now review the instruments on the Moon and the requirements they impose
on the ground apparatus.  Example performance is also presented in
Section~\ref{sub:APOLLO-Apparatus}.

\begin{figure}[tbh]
\begin{center}\includegraphics[%
  width=3in,
  keepaspectratio]{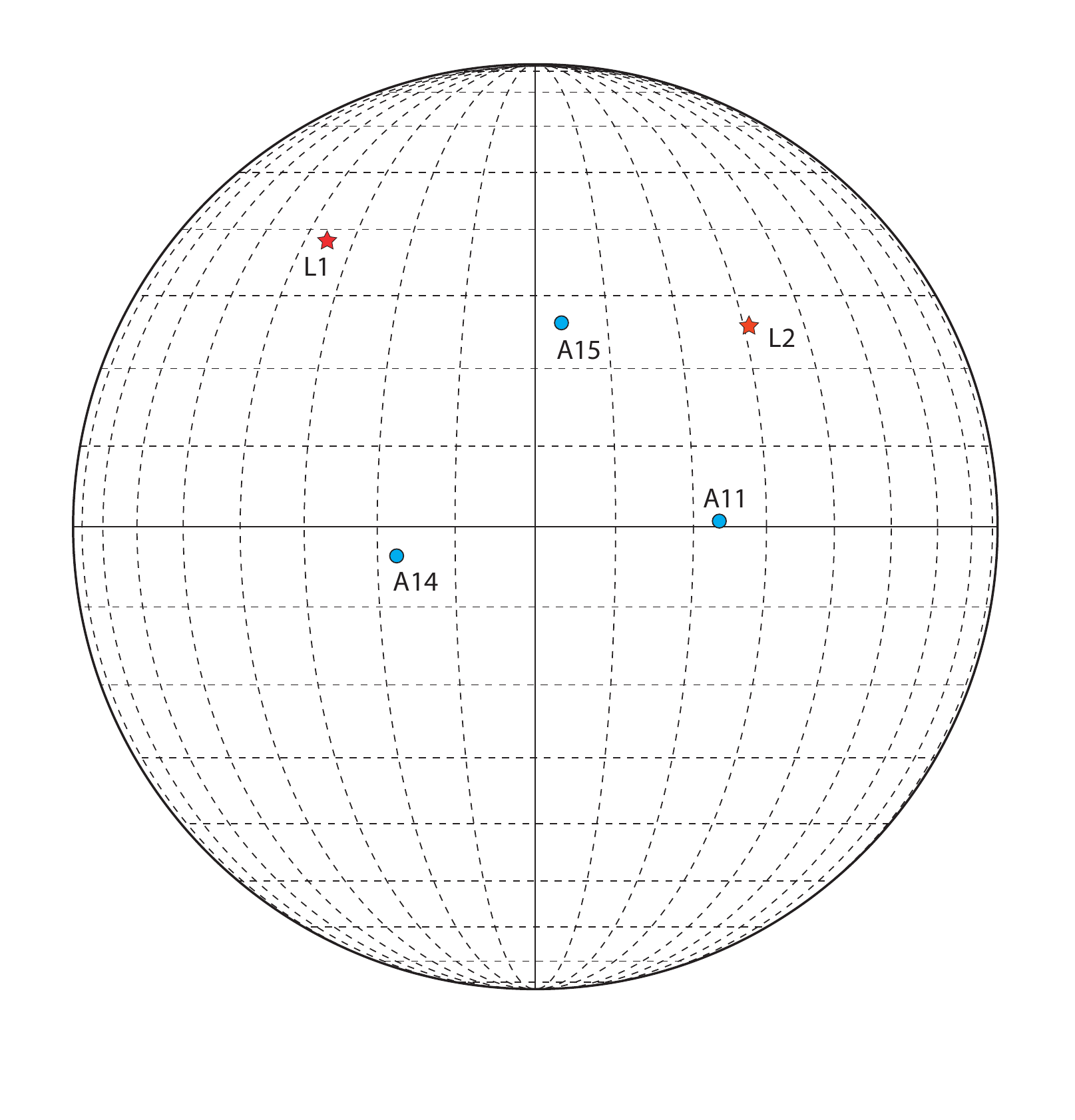}\end{center}

\caption{{\small Positions of the five reflectors on the lunar surface. ``A''
stands for Apollo, while ``L'' stands for
Lunokhod.\label{fig:locations}}}
\end{figure}

\begin{figure}[tbh]
\begin{center}
\includegraphics[height=2.5in,keepaspectratio]{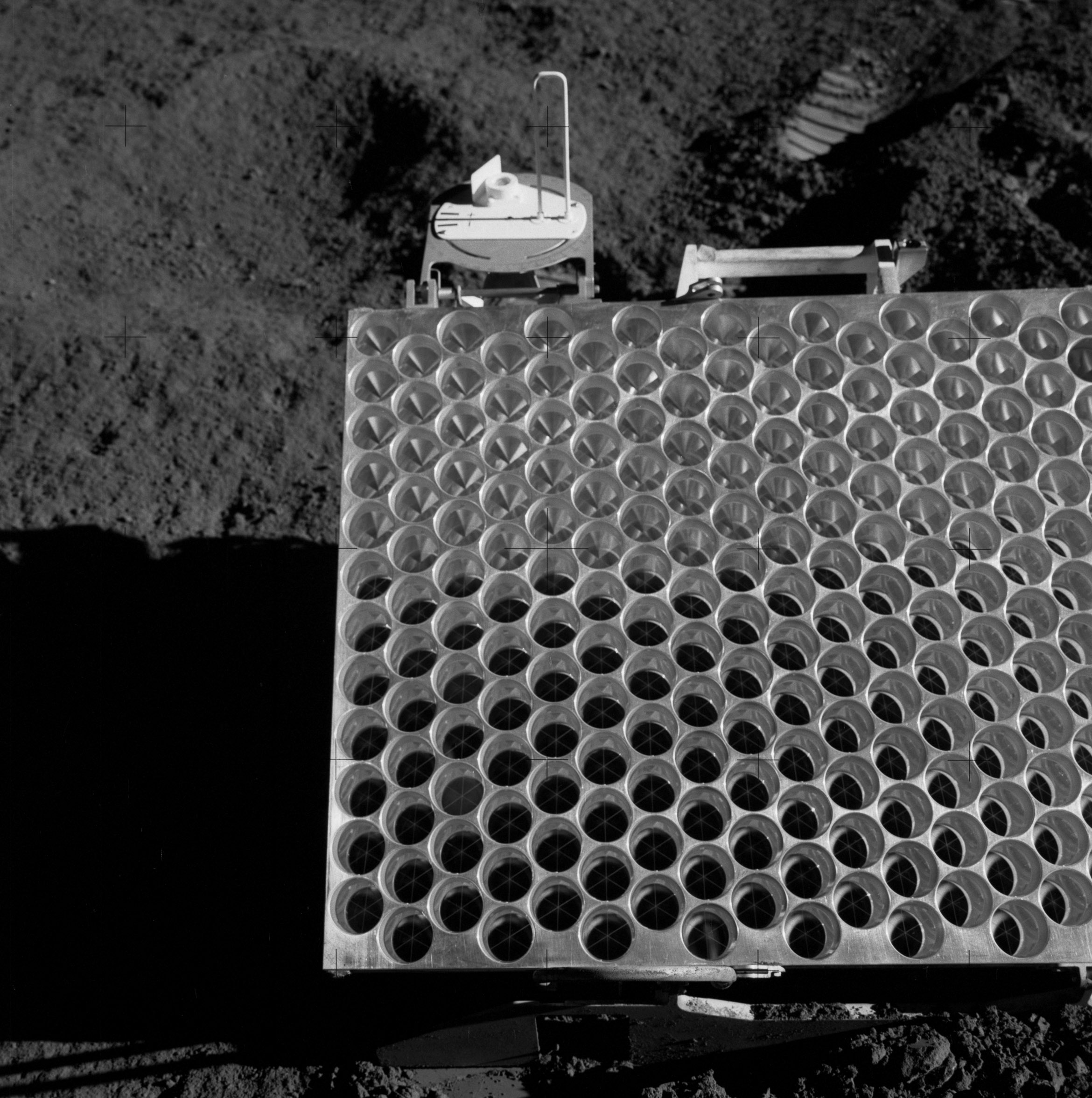}\hfill
\includegraphics[height=2.5in,keepaspectratio]{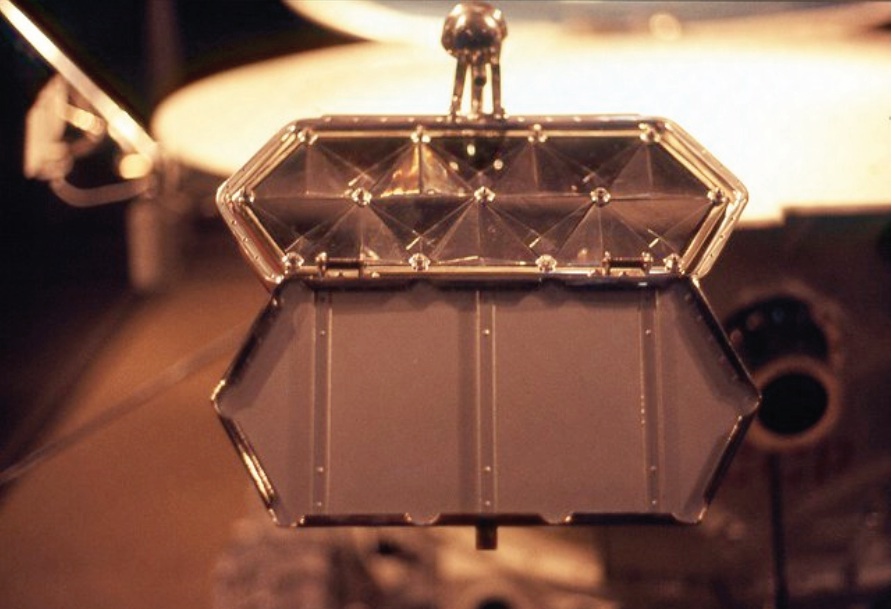}
\end{center}

\caption{{\small A portion of the Apollo~15 reflector (left; courtesy NASA), consisting of 300 corner
cube reflectors each 3.8~cm in diameter.  All Apollo reflectors are mounted
in a similar aluminum tray, except that the Apollo~11 and Apollo~14 reflectors are
100-element arrays arranged in a 10$\times$10 square pattern.  At right is
the Lunokhod reflector design (courtesy the Lavochkin
Association).\label{fig:reflectors}}}
\end{figure}

LLR relies on a total of five passive reflectors left on the surface
of the Moon roughly 40 years ago (Fig.~\ref{fig:locations}). The
Apollo arrays---landed on the Apollo~11, Apollo~14, and Apollo~15
missions---consist of, respectively, 100, 100, and 300 3.8~cm diameter
fused silica corner cube reflectors employing total internal reflection.
The Luna~17 and Luna~21 soviet missions to the Moon landed the Lunokhod~1
and Lunokhod~2 rovers, each carrying identical reflector arrays built
by the French. These arrays consist of 14 corner cubes each having
a triangular edge length of 11~cm and silvered rear surfaces. The
nominal response of the Lunokhod arrays falls between that of the
100-element and 300-element Apollo arrays.  Pictures of both types of
arrays appear in Fig.~\ref{fig:reflectors}.

\begin{figure}[tbh]
\begin{center}\includegraphics[%
  width=4in,
  keepaspectratio]{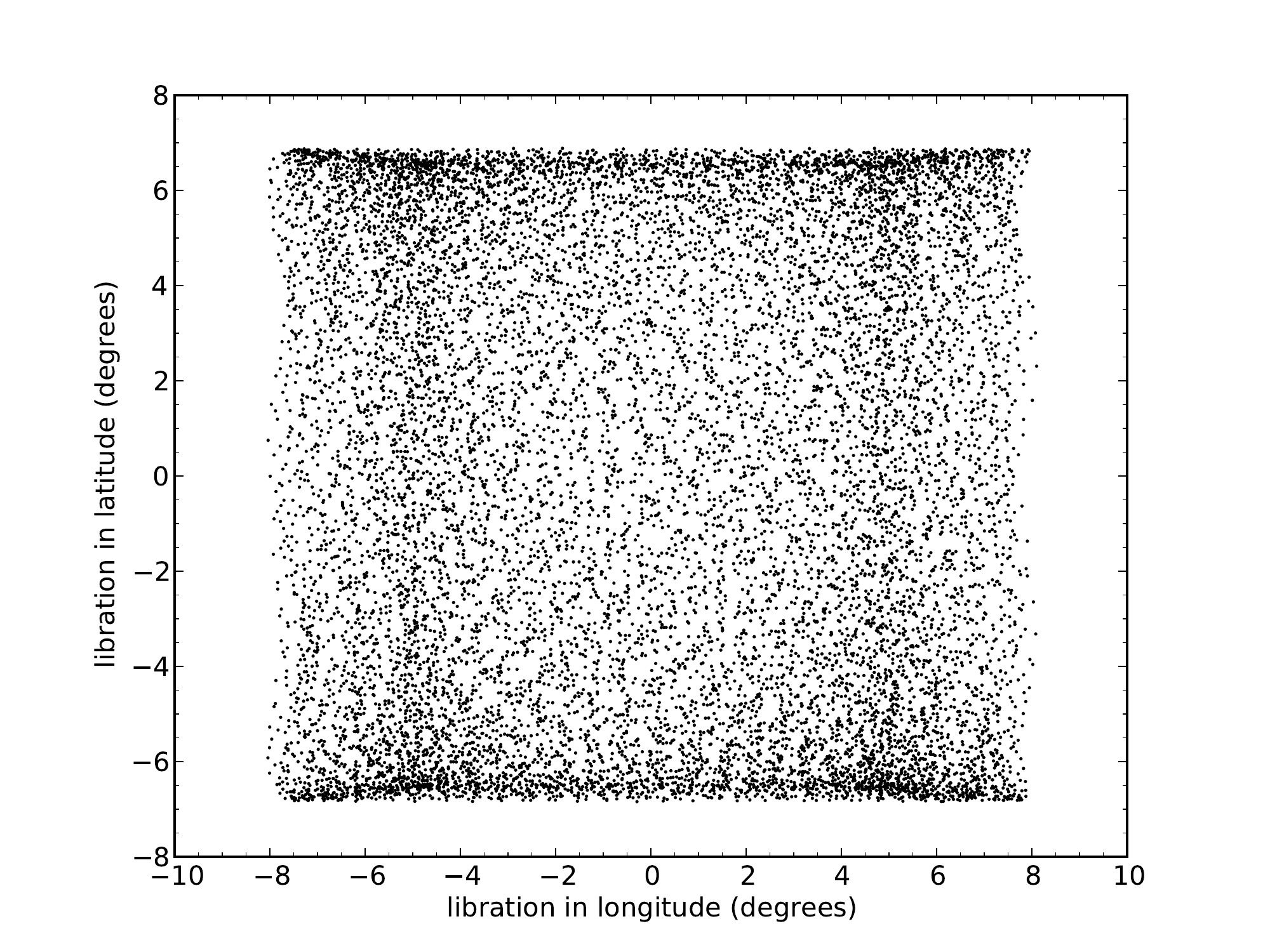}\end{center}

\caption{{\small Libration pattern for the Moon over 18.6 years, at 12~h samples.
\label{fig:librations}}}
\end{figure}

\begin{figure}[tbh]
\begin{center}\includegraphics[%
  width=4in,
  keepaspectratio]{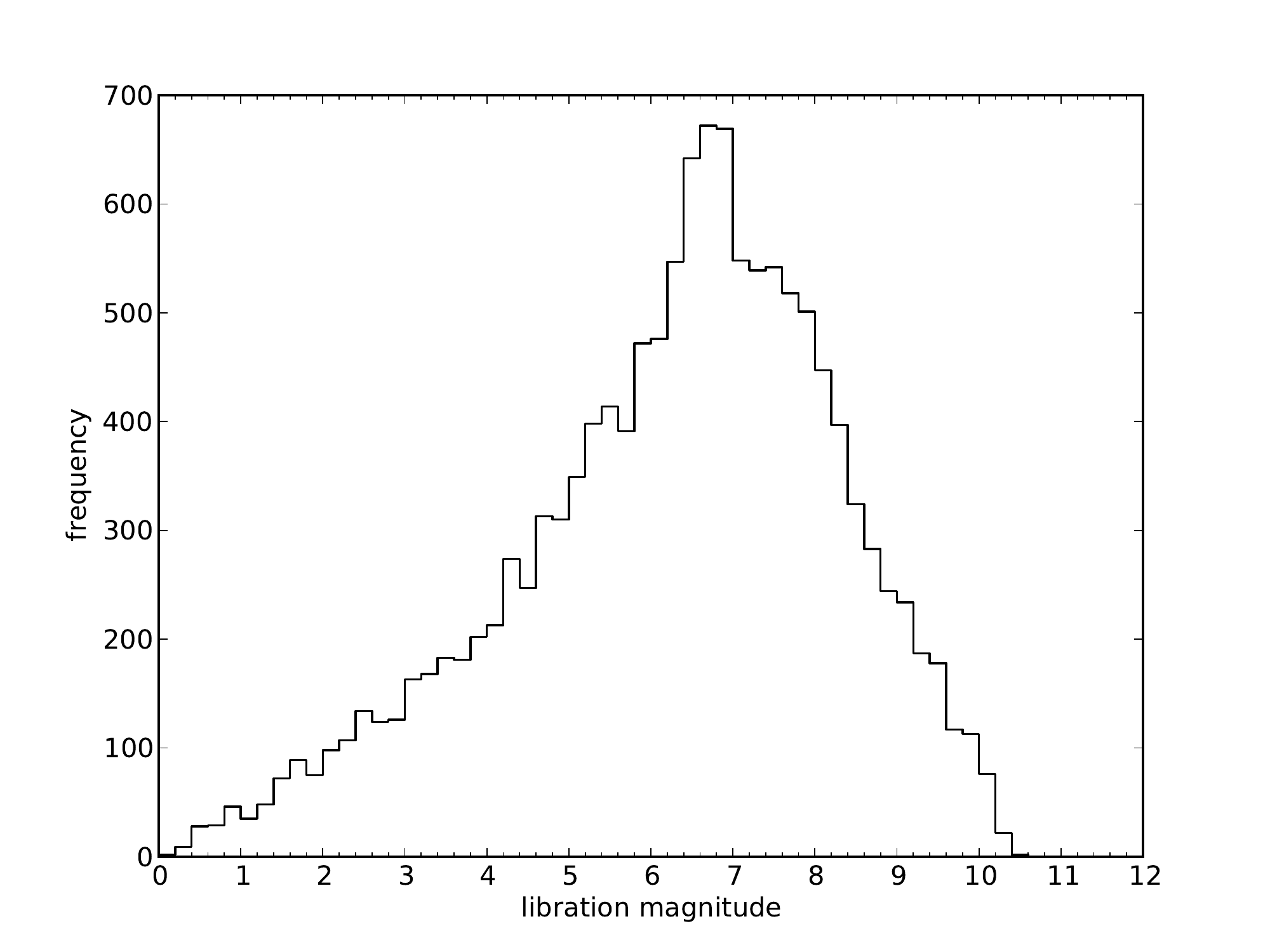}\end{center}

\caption{{\small Total libration frequency histogram, ignoring topocentric
contributions.\label{fig:libration-histogram}}}
\end{figure}

Lunar libration changes the apparent tilt of the reflectors with respect
to the Earth-Moon line of sight, seen in Fig.~\ref{fig:librations}
filling out a rectangle spanning $\pm8.1^{\circ}$ in longitude and
$\pm6.9^{\circ}$ in latitude---not including the Earth topocentric
correction, which can modify effective libration by as much as a degree.
The median total libration is $6.5^{\circ}$, and can be in excess
of $10^{\circ}$ (Fig.~\ref{fig:libration-histogram}), again ignoring
topocentric considerations. As a result, the return pulse acquires
a temporal spread due to the fact that some corner cube reflectors
are closer to the observer, while others are farther. For the Apollo~15
array, the full-width at half-maximum (FWHM) can approach 1~ns (150~mm
one-way), or a root-mean-square (RMS) in excess of 300~ps. A typical
case of 200~ps RMS corresponds to 30~mm of one-way distance, and
thus requires 900 measurements (photons) in order to achieve a statistical
uncertainty in the neighborhood of 1~mm. Thus precision LLR demands
hundreds or thousands of photons in order to overcome the libration/reflector-imposed
measurement uncertainty.

\begin{figure}[tbh]
\begin{center}\includegraphics[%
  width=4in,
  keepaspectratio]{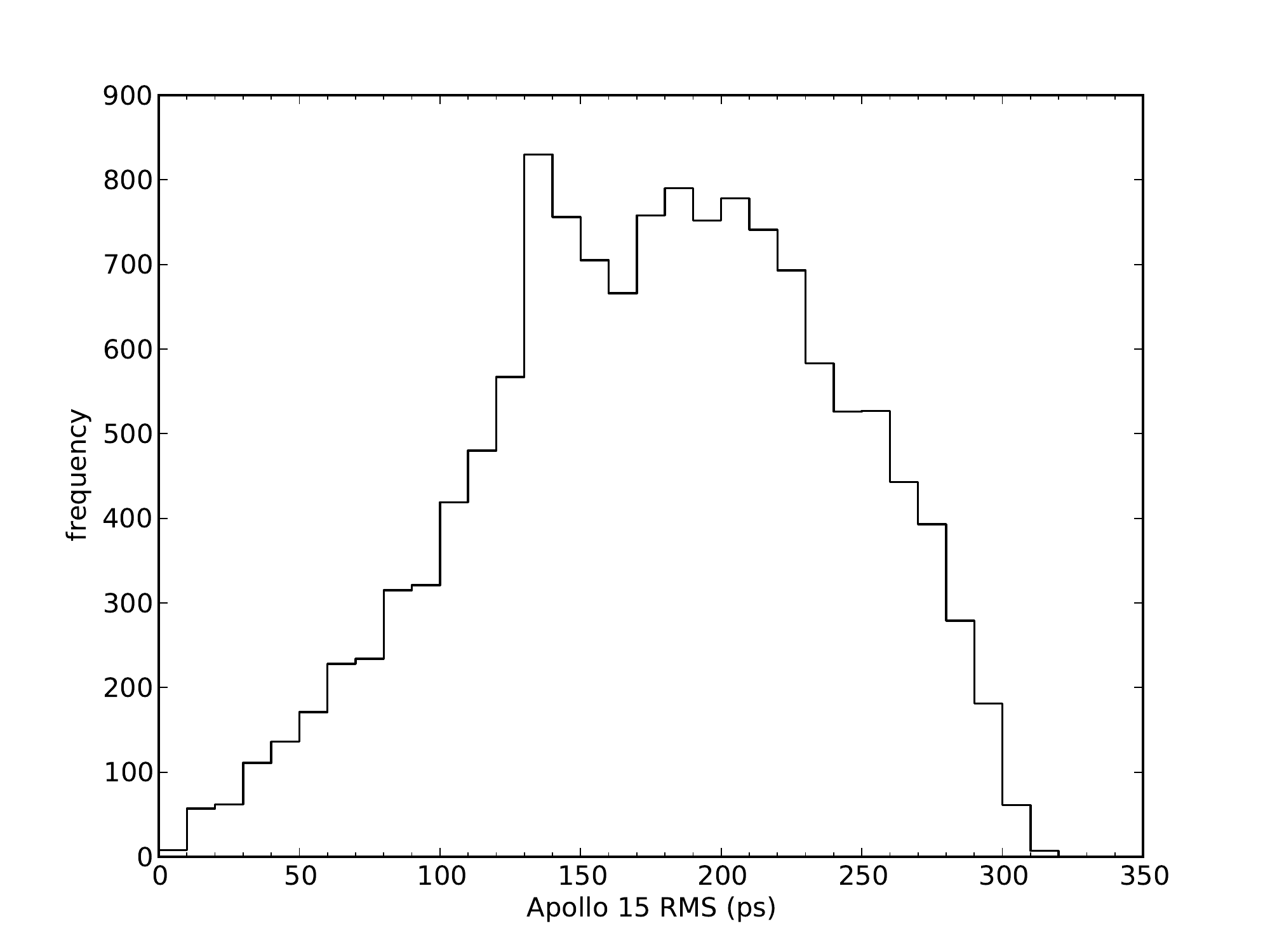}\end{center}

\caption{{\small Impact of librations on the root-mean-square temporal variation contributed
by the Apollo~15 reflector.\label{fig:A15-RMS}}}
\end{figure}

The signal loss in the two-way laser link is staggeringly high, generally
amounting to a loss factor in the neighborhood of $10^{18}$. Beam
divergence on both the up-leg and down-leg result in a signal strength
that depends on the inverse-fourth power of distance. 

The up-leg divergence is limited by atmospheric turbulence (seeing),
so that one may not generally expect better than about 1~arcsec (5~$\mu$rad)
divergence, translating to about $4\times10^{-8}$ throughput onto
the smaller Apollo~11 and 14 arrays. The down-leg divergence is set
by diffraction from the corner cubes. Total internal reflection corner
cubes like those used for Apollo produce a central irradiance that
is 0.15 times that of a simplistic top-hat illumination pattern of
angular diameter $\lambda/D$, where $\lambda$ is the wavelength
and $D$ is the corner cube diameter. The result is that a 1~m circular
aperture on Earth receives about $5\times10^{-9}$ of the flux incident
on the reflector at a wavelength of 532~nm. The net throughput is
then $\sim2\times10^{-16}$ for a 1~m aperture, scaling as $a^{2}/\lambda^{2}$,
where $a$ is the aperture diameter. Multiplying by typical optical
system and atmospheric throughputs (traversed twice), together with
filter transmissions and detection efficiencies, total throughput
tends to be in the range of $10^{-18}$. An energetic pulse of laser
light having a pulse width in the neighborhood of 100~ps might be
100~mJ, containing $3\times10^{17}$ photons at green wavelengths.
The result is that LLR invariably operates in the single-photon detection
regime. 

The ground apparatus therefore benefits from having:

\begin{itemize}
\item a powerful laser, typically a few Watts, with substantial pulse energy;
\item sub-arcsecond intrinsic divergence, meaning a launch beam diameter
exceeding 10~cm;
\item sub-arcsecond pointing and tracking capability;
\item a large collection aperture for the return;
\item aggressive filtering in the temporal, spatial, and wavelength domains.
\end{itemize}
Additionally, the laser pulsewidth should be $<100$~ps so that it
does not compete with the reflector-induced spread in the overall
error budget. Many of the requirements are naturally satisfied by
using an astronomical telescope as both the launch and receive instrument
in a so-called mono-static arrangement. For perspective, a 1~W laser
limited to 1~arcsec divergence by the atmosphere will return approximately
$10^{-16}$~W~m$^{-2}$ to the ground, translating to a 19$^{\mathrm{th}}$
magnitude source viewed through a 100~nm wide broadband filter. Meanwhile,
the full moon reaches $-13$~mag, or about $10^{13}$ times brighter.
More fairly, the surface brightness of the full moon is about 3~mag~arcsec$^{-2}$,
so that an aperture (spatial filter) spanning four square arcseconds
admits $10^{7}$ times more background than signal, or a signal-to-background
ratio (SBR) around $10^{-7}$. A wavelength filter having a 1~nm
passband increases the SBR to $\sim10^{-5}$, and temporal filtering
at the 1~ns level compared to a 50~ms repetition rate (20~Hz) carries
a factor of $5\times10^{7}$ for a net SBR$\approx500$. Clearly,
the temporal filter is the most effective of the three. Missing from
this discussion is the apparent degradation of the lunar reflector
response \cite{dust}, reduced by a factor of ten across the board,
and an additional factor of ten at full moon (discussed in Section~\ref{sub:Reflector-Degradation}).

\begin{figure}[tbh]
\begin{center}\includegraphics[%
  width=6in,
  keepaspectratio]{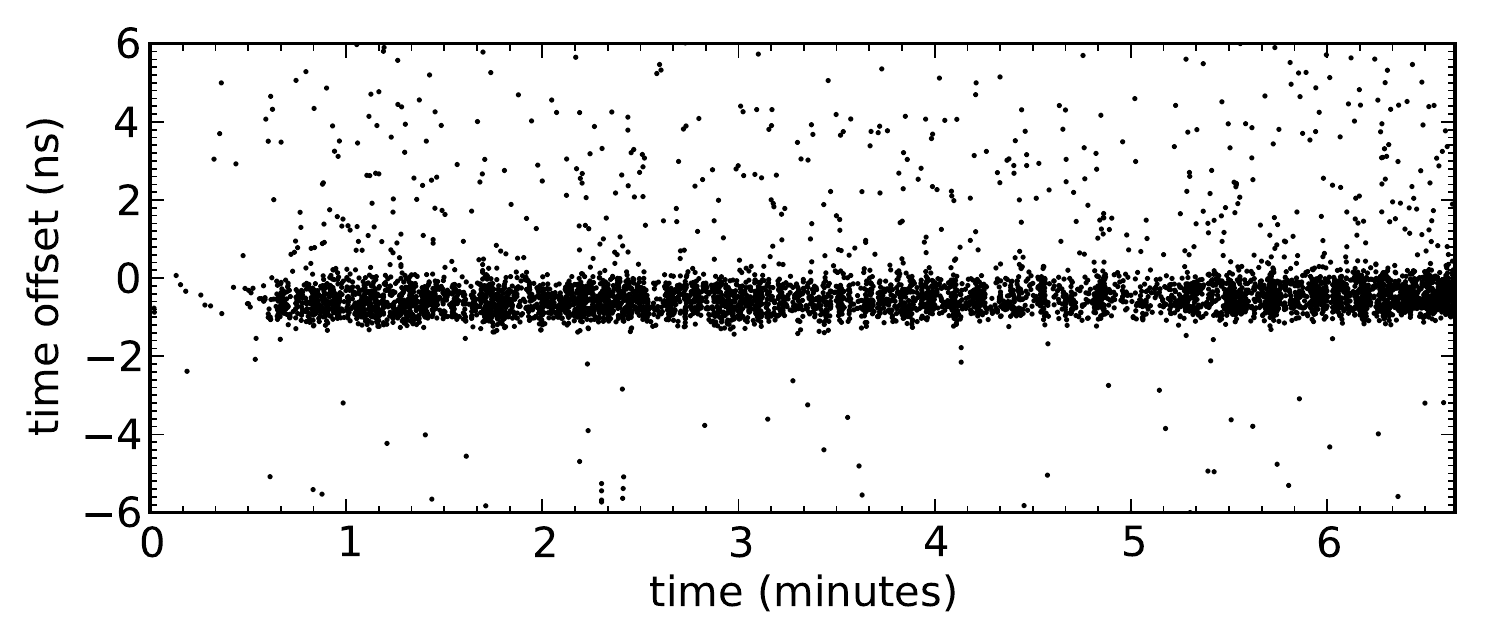}\end{center}

\caption{{\small 8000-shot measurement to Apollo 15 on 2010 March 23, showing a 12~ns
portion of the 100~ns return window. This is the first run of the
night, demonstrating a typical fast acquisition and optimization of
the signal. Background photons and detector dark events are scattered
below (before) the lunar return. A diffusion process in the detector
contributes to a ``tail'' of late responses after the lunar return.
It is typical for the timing of the return to differ from an approximate
prediction by $\sim1$~ns, as seen here. The temporal thickness of
the Apollo~15 return is due to the finite size of the slightly-tilted
reflector, as seen more clearly in
Fig.~\ref{fig:A15-hists}.\label{fig:A15-run}}}
\end{figure}

\begin{figure}[tbh]
\begin{center}\includegraphics[%
  width=6in,
  keepaspectratio]{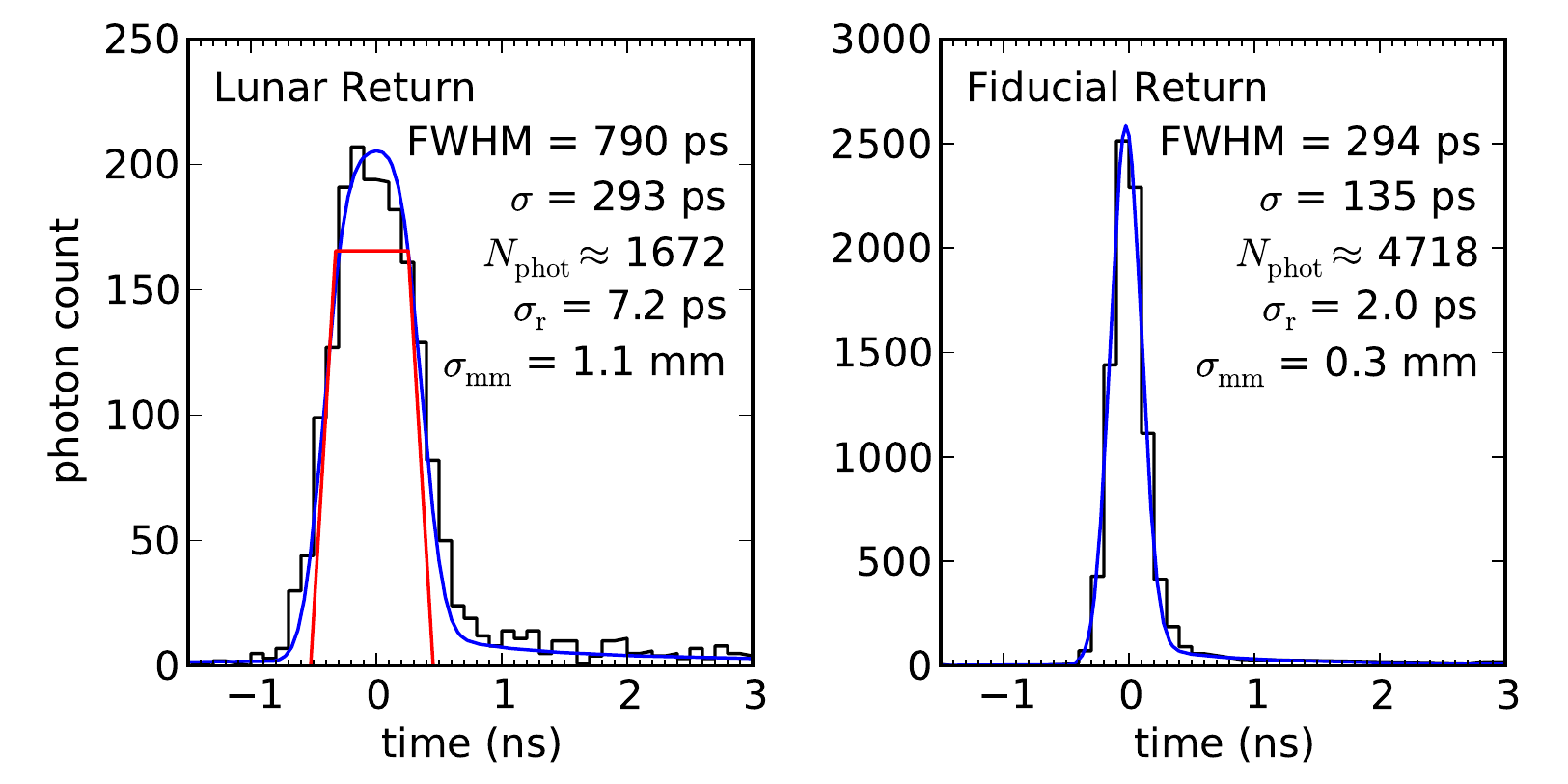}\end{center}

\caption{{\small Histogram of the lunar return from Fig.~\ref{fig:A15-run} (left),
along with a fiducial measurement reflecting the system performance
(right). A functional fit to the fiducial return (from a local corner
cube) is convolved with the trapezoidal shape resulting from the total
libration---lunar plus Earth topocentric correction---at the time
of observation to make the fit line for the lunar return. The FWHM
and RMS of each distribution is given, along with the number of contributing
photons and a measure of centroiding ability given by statistical
reduction of the RMS by $\sqrt{N}$, expressed in picoseconds and
millimeters. \label{fig:A15-hists}}}
\end{figure}

An example return from Apollo~15 is shown in Fig.~\ref{fig:A15-run},
demonstrating the effect of libration as well as the high SBR recently
achieved in LLR measurements. A histogram of the same data is presented
in Fig.~\ref{fig:A15-hists}, in which it is seen that the temporal
width of the lunar return is almost completely determined by the libration-induced
reflector spread. Details on the apparatus used to acquire these data
can be found in Section~\ref{sub:APOLLO-Apparatus}.

\subsection{Fundamental Measurement and World Lines\label{sub:Fundamental-Measurement-and}}

It is important to understand the fundamental measurement performed by LLR.
After a brief description in this section, a more detailed treatment may be
found in Section~\ref{sub:Model-content/construction}.  A stable clock
provides a frequency reference by which time intervals may be precisely
measured. Time transfer techniques---usually via the Global Positioning
System (GPS)---guarantee long-term frequency stability and synchronization
with atomic clock ensembles around the globe.

The core measurement consists of recording two event times---corresponding
to photon launch and detection---against the local clock. The launch
time is generally gauged by measuring the return from a local corner
cube mounted at the telescope exit aperture, heavily attenuated to
the single-photon level so that the same detector and timing system
may be used for both the local (fiducial) and remote (lunar) photon
returns. A constant offset between the Earth-fixed axis intersection
of the telescope and fiducial corner cube is added to each range measurement.
In order to achieve millimeter-level precision, the \emph{absolute}
time only needs to be accurate at the microsecond level (Earth rotation
modifies the Earth-Moon distance by $\sim0.4$~mm in 1~$\mu$s),
while the \emph{relative} time must be understood at the few-picosecond
level. The latter requirement translates to frequency stability at
the $10^{-12}$ level over the course of a few seconds.

\begin{figure}[tbh]
\begin{center}\includegraphics[%
  width=2.5in,
  keepaspectratio]{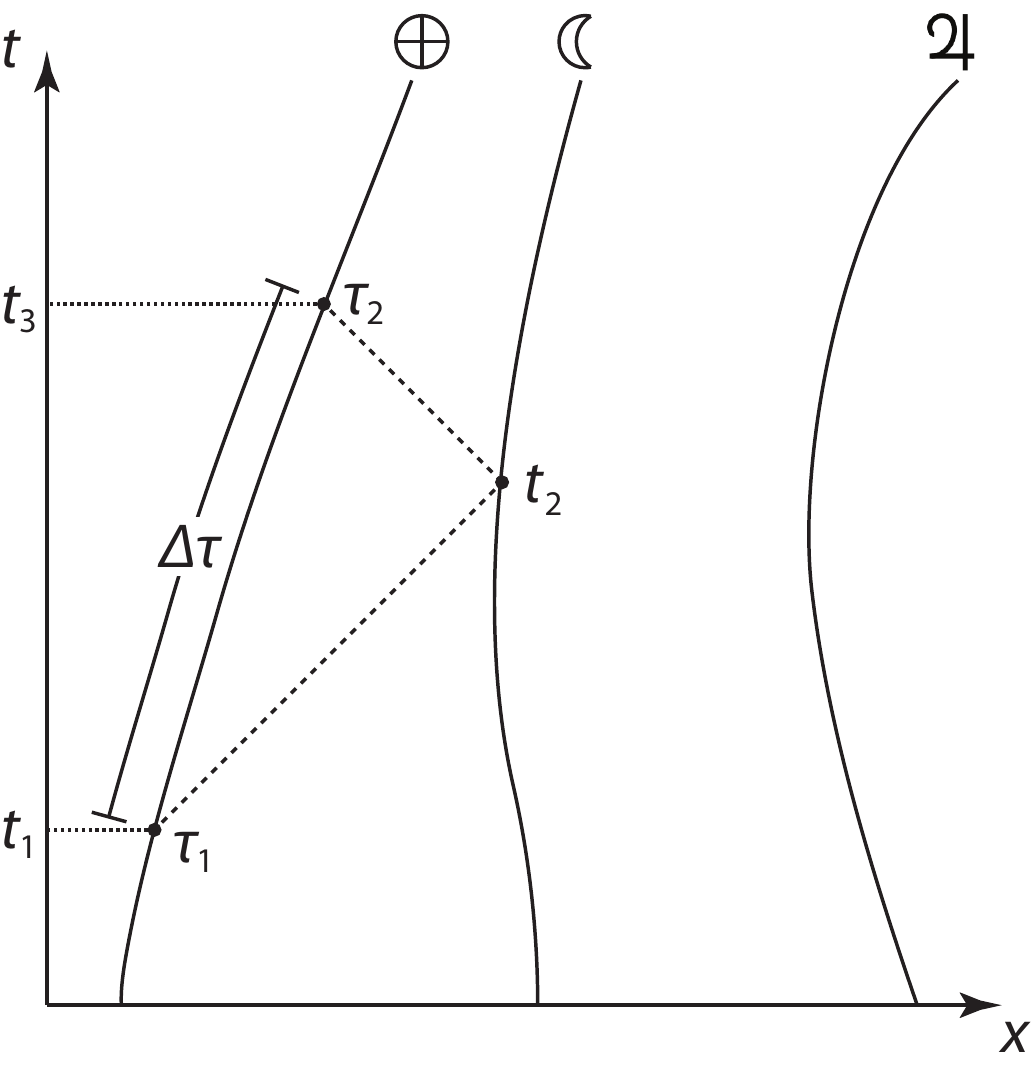}\end{center}

\caption{{\small Schematic of fundamental LLR measurement. The Earth and Moon move
through the SSB frame, influencing each other and being influenced
by other solar system bodies (e.g., the Sun, Jupiter). A light pulse
travels a null geodesic from the Earth to the Moon and back, while
the Earth clock ticks off a proper time, $\Delta\tau=\tau_{2}-\tau_{1}$.
The event times are transformed into the SSB frame ($t_{1}$ and $t_{3}$)
and the entire solar system model is adjusted (including the bounce
time, $t_{2}$) until the various world lines are made to match the
entire set of LLR measurements.\label{fig:world-lines}}}
\end{figure}

Because all massive solar system bodies influence the Earth-Moon range, the
analysis is most conveniently performed in the solar system barycenter
(SSB) frame. The measured times are transformed into SSB coordinates using
standard time transformation techniques---as outlined, for example, by
Moyer (1981) \cite{moyer}---primarily consisting of adjustments to account
for velocity-induced time dilation and gravitational redshift resulting
from the solar potential. Accounting for body figures and rotation of Earth
and Moon (and the solar $J_{2}$), a fully relativistic
(Einstein-Infeld-Hoffmann: EIH) equation of motion is numerically
integrated, varying input parameters---chiefly initial conditions---to
search for dynamical world-lines (Fig.~\ref{fig:world-lines}) of the
relevant bodies that satisfy the round-trip light propagation measurements
in the SSB frame. Parameters in the relativistic model allowing departures
from the specific prescription of general relativity (GR) indicate
deviations from GR. The technique is more fully described in Williams et
al. (1996) \cite{jgw96} and M\" uller et al. (2008) \cite{goed-and-rel}.
Simultaneous numerical integration of lunar rotation is critical for
computing consistent dynamics, and will be treated further in
Section~\ref{sub:The-Lunar-Interior}.

\section{Science from LLR\label{sec:Science-from-LLR}}

Here we present a more comprehensive account of the science delivered by
LLR than was introduced in Section~\ref{sub:Current-Science-Results}.
First, we look at gravitation, including the equivalence principle,
time-rate-of-change of Newton's $G$, gravitomagnetism, geodetic precession,
and the inverse-square law.  Next, we briefly address the lunar interior
and geophysical concerns.

\subsection{Relativity and Gravity\label{sub:Relativity/Gravity}}

The concordance of astrophysical measurements in the last 15 years---the
anisotropy scale of the cosmic microwave background \cite{boomerang,maxima,boomerang2,dasi,wmap},
the distance measurements of Type~Ia supernovae \cite{sn1,sn2},
the gravitational behaviors of galactic superclusters \cite{cluster},
and the power spectrum of large-scale structure \cite{bao}---point
to the surprising conclusion that the expansion of the universe is
\emph{accelerating}, implying some form of a \emph{fundamentally new}
gravitational phenomenon. The cosmological acceleration could be due
to a scalar field that produces effects similar to those associated
with the ``cosmological constant,'' originally introduced into
the relativistic field equations by Einstein. A scalar field would
likely couple to the gravitational field in such a way as to produce
a departure from the equivalence principle (EP) \cite{damourwhy},
and would introduce time variations in the fundamental coupling constants
of nature \cite{damour}. EP and $\dot{G}$ tests therefore have discovery
potential with a very broad reach, and in fact provide some of the
most sensitive low-energy probes for new physics. In light of recent
discoveries, it is important that scientific inquiry is not restricted
to current theoretical expectations, but rather that every available
avenue for testing the nature of gravity is examined.

The lunar orbit offers a pristine laboratory for testing gravity,
as non-gravitational effects on the orbit begin to show up only at
the millimeter level. Moreover, the Moon is far enough from the Earth
to be dominated by solar gravity, so that the Earth and Moon may each
be considered to be in \emph{solar} orbits. This fact makes the Earth-Moon-Sun
system useful as a probe of the equivalence principle (and other relativistic
phenomena) at scales of 1~AU---extending the baseline against which
to compare the raw measurement precision.

We highlight here some of the contributions to gravitational physics
from LLR. The list is not complete, but provides a sense of the cornerstone
capabilities. Most of these science results are based on modeling
that currently produces post-fit residuals of measured data in the
neighborhood of 2~cm, so that millimeter-quality data could in principle
improve current limits by an order of magnitude given commensurate
improvements in modeling (see Section~\ref{sec:The-Modeling-Challenge}).

\subsubsection{Equivalence Principle\label{sub:Equivalence-Principle}}

The simplest prediction of Einstein's equivalence principle---the
universality of free-fall---is one of the most precisely tested principles
in all of physics. Yet there are strong motivations for extending
the tests and pushing their precisions even higher. The EP can be
decomposed into two key forms. The weak form of the EP (WEP) applies
to the gravitational properties of all forms of mass-energy except
for gravity, while the strong EP (SEP) extends the WEP to include
gravity itself. 

The Earth-Moon-Sun system is currently the \emph{best available} probe
of the SEP, first pointed out by Nordtvedt \cite{nord1,nord2,nord3}.
From the vantage point of the EP, the Earth and Moon are test bodies
that differ in two important ways. First, the Earth's mass has a fractional
contribution from gravitational self-energy ($4.6\times10^{-10}$
) that is about 20 times greater than the corresponding measure for
the Moon---allowing LLR to test the SEP. Second, the Earth has a massive
iron-nickel core while the Moon does not---making LLR sensitive to
a WEP violation as well. Laboratory EP tests of Earth-like and Moon-like
objects falling toward the sun can be used to distinguish between
an SEP and a WEP violation \cite{npl}. 

LLR tests the SEP by measuring the difference in the accelerations of the
Earth and Moon toward the Sun. In the presence of a differential
acceleration, the orbit of the Moon---from our perspective on the
Earth---would appear to be displaced, or \emph{polarized}, toward or away
from the Sun. The range signal would take the form\[ \Delta
r\cong13\eta\cos D\,\,\mathrm{meters}\]
 where $D=(\omega-\Omega)t$ is the lunar orbit's synodic
phase\footnote{The
synodic phase of the Moon describes its angle with respect to the Earth-Sun
line, thus referring to the familiar illumination cycle of lunar phases.}
having a period of 29.53 days, with $D=0$ corresponding to new moon
\cite{nord}.  The parameter $\eta$ is a theory-dependent dimensionless
coefficient \emph{sensitive to almost every post-Newtonian feature of the
theory}.  Although $\eta$ vanishes in general relativity, it generally does
not in alternative theories. But independent of any theory, this test of
the SEP addresses a very basic and important question---what is the weight
of gravity itself? It tests a crucial non-linear property of gravity: how
gravity produces energy that itself gravitates.

The metric models by Damour and Nordtvedt \cite{dn} describe a relaxation
of scalar field strength that today would produce SEP differential
accelerations between $5\times10^{-17}$ and $10^{-13}$. The present
limit on differential acceleration is $\Delta a/a\approx\pm1.3\times10^{-13}$
\cite{jgw-ep,jgw-cqg-2012,muller-cqg-2012}, corresponding to a test
of the SEP at the level of $|\eta|<3\times10^{-4}$, given the self-energy
fraction of the Earth. Millimeter-quality ranging stands to improve
sensitivity of the SEP test by one order of magnitude, measuring $\Delta a/a$
to a precision of $\pm10^{-14}$ and reaching into the theoretically
motivated range indicated above. The closest competitor comes from
pulsars. A composite of 27 pulsars places a 1-$\sigma$ limit on the
SEP of $2.3\times10^{-3}$---roughly an order of magnitude shy of
the LLR result \cite{gonzalez,freire-2012}.

\subsubsection{Time-rate-of-change of $G$\label{sub:Time-rate-of-change-of-G}}

A secular change in the gravitational constant, $G$, would produce
secular changes in the lunar mean distance and the orbital period
(Kepler's third law), as well as in the angular rate of the Earth
about the Sun. While the orbital radius change results in a range
signal that varies linearly in time, the change in orbital period
leads to a quadratic evolution of the Moon's mean anomaly (phase).
It is this quadratic dependence that most powerfully constrains $\dot{G}$.
Here, the long time span of LLR measurements becomes important, limiting
$\dot{G}/G$ at the impressive level of $7\times10^{-13}\,\mathrm{yr}^{-1}$
\cite{mueller-gdot} and $9\times10^{-13}\,\mathrm{yr}^{-1}$ \cite{jgw-2004}---the
best available experimental results. 

Recently Steinhardt and Wesley examined the constraints that observations
and experiment place on a broad class of theories that attempt to
explain dark energy in the context of extra-dimensions \cite{steinhardt}.
They find that if current constraints on \emph{both} $\dot{G}$ \emph{and}
the value and rate-of-change of the equation-of-state parameter, $w$,
improve by a factor of two, such ideas could be ruled out at the 3$\sigma$
level. In their analysis, Steinhardt and Wesley use a 1994 pulsar
timing limit for $\dot{G}$ of $5\times10^{-12}$. LLR already exceeds
this limit by an order of magnitude. A recent work questions the validity
of several previous published pulsar limits on $\dot{G/G}$---including
the 1994 result---finding instead trustworthy limits in the neighborhood
of $20\times10^{-12}$ per year \cite{verbiest}. However, a new report
puts forth a limit at $1.6\times10^{-12}$ per year, approaching levels
tested by LLR \cite{freire-gdot}.

\subsubsection{Gravitomagnetism, Geodetic Precession, and other PPN Tests\label{sub:Gravmag-Geodetic-etc}}

LLR tests a number of basic relativistic phenomenologies---independent
of whether gravity is described by a metric theory. These phenomena
include gravitomagnetism, geodetic precession, and the consequences
of preferred frames. Many such phenomena can be cast into the Parameterized
Post-Newtonian (PPN) framework \cite{ppn,will}: a generalized metric
description of gravity for which general relativity is a special case.
The most prominent PPN parameters are $\gamma$, describing the amount
of curvature produced per unit mass, and $\beta$, describing the
non-linearity of gravity. Both of these are unity in general relativity.
The best constraint on $\gamma$ comes from Doppler ranging to Cassini:
$|\gamma-1|<2.3\times10^{-5}$ \cite{cassini}. $\beta$ is best constrained
by LLR tests of the SEP via the identity: $\eta\equiv4\beta-\gamma-3$.
Combining the Cassini result for $\gamma$ with the LLR result for
$\eta$ yields $|\beta-1|<1\times10^{-4}$ \cite{jgw-ep}. 

Preferred frame effects, such as those codified by PPN parameters
$\alpha_{1}$ and $\alpha_{2}$, are also tested by LLR, currently
at the level of $9\times10^{-5}$ and $2\times10^{-5}$, respectively
\cite{mueller-gdot,mueller-a1}, although Nordtvedt obtains a $10^{-7}$
limit on $\alpha_{2}$ based on the long-term orientation of the spin
axis of the Sun \cite{nordtvedt-a2}.

Gravitomagnetism is a generic consequence of any mass in motion. As
the Earth orbits the Sun, its gravitomagnetic field exerts a Lorentz
force on the Moon. Eliminating the gravitomagnetic term from the EIH
equations of motion would result in experimentally-absent six-meter-amplitude
periodic disturbances at both synodic and twice-synodic frequencies
\cite{gravmag-prl}. LLR constrains gravitomagnetism---the root of
``frame dragging''---to $< 0.2$\% precision, as confirmed by a covariant
analysis \cite{soffel}. Within the PPN context, gravitomagnetism
reduces to non-competitive checks on parameters $\gamma$ and $\alpha_{1}$.
By comparison, the Gravity Probe-B experiment obtained a final precision
on the gravitomagnetic effect of 19\% \cite{gpb-final}, and laser
ranging to the LAGEOS satellites produce results in the 5--40\% range
\cite{ciufolini-lageos,iorio}. Pulsars may soon contribute gravitomagnetic
limits as well, although no pulsar results have been published to
date.

Gravitomagnetism, at its core, is a frame-dependent phenomenology.
As such, the assertion that LLR is sensitive to gravitomagnetism has
been questioned from the point of view that one may nullify the effect
by performing LLR analysis in an Earth-centered frame \cite{kopeikin-comment,ciufolini}.
Leaving aside complications arising from the fact that the resulting
frame is non-inertial (not asymptotically flat), the attempt to separate
gravitomagnetism into ``intrinsic'' and ``gauge-dependent''
varieties is, in the author's view, as specious as it would be for
the magnetic field of electromagnetism: there are not two physically
distinct flavors of magnetic fields. A similar argument could be made
that performing analysis of LAGEOS or Gravity Probe-B measurements
of Lense-Thirring or Schiff precessions, respectively, in a frame
rotating with the Earth would likewise eliminate the source of gravitomagnetism.
Obviously other frame-dependent phenomenologies should intercede to
produce the same observational result, but this merely amplifies the
notion of gravitomagnetism as part of the frame-transformation package.
Keeping in mind that converting measurements into the SSB frame for
LLR analysis reduces to a straightforward matter of time transformation,
as summarized in Section~\ref{sub:Fundamental-Measurement-and},
the lack of anomalous gravitomagnetic signatures when evaluating LLR
data in the context of the EIH equations of motion essentially stands
as confirmation that gravitomagnetism plays its expected role in frame
transformation \cite{gravmag-prose}. The appearance of the PPN preferred
frame parameter, $\alpha_{1}$, in the coefficient for the gravitomagnetic
term in the equations of motion further clarifies this association.

Geodetic precession can be understood as the effect of parallel transport
of a fixed direction---as manifested by a gyroscope or orbital axis,
for instance---around the curved space surrounding a central body.
The curvature results in a migration of the axis direction relative
to the background inertial space upon completing an orbit. One revolution
around the Sun at the distance of the Earth produces a directional
offset of 19.2 milliarcseconds. This precession rate in the orientation
of the lunar orbit is confirmed by LLR \cite{jgw96,jgw-2004}, presently
at the 0.3\% level, corresponding to about 60~$\mu$as~yr$^{-1}$
\cite{jog-rev}. Geodetic precession is primarily another measure
of PPN $\gamma$. The Gravity Probe-B mission finalized a result on
geodetic precession at the 0.28\% level. Pulsar timing has thus far
confirmed geodetic precession at the 13\% level \cite{breton}.

\subsubsection{Inverse Square Law, Extra Dimensions, and other Frontiers\label{sub:Inverse-Square-Law,}}

Any deviation from the Newtonian $1/r^{2}$ force law produces a precession
of orbital perigee. LLR's measurement of any anomalous precession
rate of the lunar orbit limits the strength of Yukawa-like long-range
forces with ranges comparable to the $\sim10^{8}$~m scale of the
lunar orbit to $<2\times10^{-11}$ times the strength of gravity
\cite{muller-isq}.
This is the strongest available constraint on the inverse square law
at any length scale \cite{ega-isl}.

Measurement of the precession rate can also probe a recent idea (called
DGP gravity) in which the accelerated expansion of the universe arises
not from a non-zero cosmological constant but rather from a long-range
modification of the gravitational coupling, brought about by higher-dimensional
effects \cite{lue,dvalimoon,extradim}. Even though the lunar orbit
is far smaller than the Gigaparsec length-scale characteristic of
the anomalous coupling, there would be a measurable signature of this
new physics, manifesting itself as an anomalous precession rate at
about $5\,\mu$as~yr$^{-1}$---roughly a factor of 10 below current
LLR limits, and potentially reachable by millimeter-quality LLR.

Another example of new tests that LLR can perform is represented by
the Standard Model Extension (SME), in which Lorentz-violating terms
are introduced into the Standard Model of physics in order to generalize
it \cite{sme}. Expressed in the gravitational sector \cite{sme-grav},
the SME exerts some influence on the lunar orbit \cite{bailey-SME}.
Accordingly, LLR has been used to place constraints on the relevant
SME parameters\cite{battat-SME}.

Offering a high-precision measurement of a clean dynamical system,
LLR constitutes a comprehensive check on gravitational phenomenology.
As such, we can expect that LLR will continue to exhibit sensitivity
to future theories that challenge the foundation of general relativity.

\subsection{Lunar and Earth Physics\label{sub:Lunar-and-Earth}}

While tests of gravitation constitute a compelling suite of scientific
motivations for pursuing LLR, details of the Earth and Moon also influence
the range measurement and as such open lines of inquiry into the natures
of these bodies. A more detailed treatment of these aspects appears
in another review \cite{jog-rev}, and are here briefly summarized.

\subsubsection{The Lunar Interior\label{sub:The-Lunar-Interior}}

By virtue of the fact that the Moon's orientation, orbit, and tidal
deformation are influenced by the interaction of its internal mass
structure with torques and forces imposed upon it, LLR data can expose
properties of the lunar interior otherwise unavailable.

\begin{figure}[tbh]
\begin{center}\includegraphics[%
  width=4in,
  keepaspectratio]{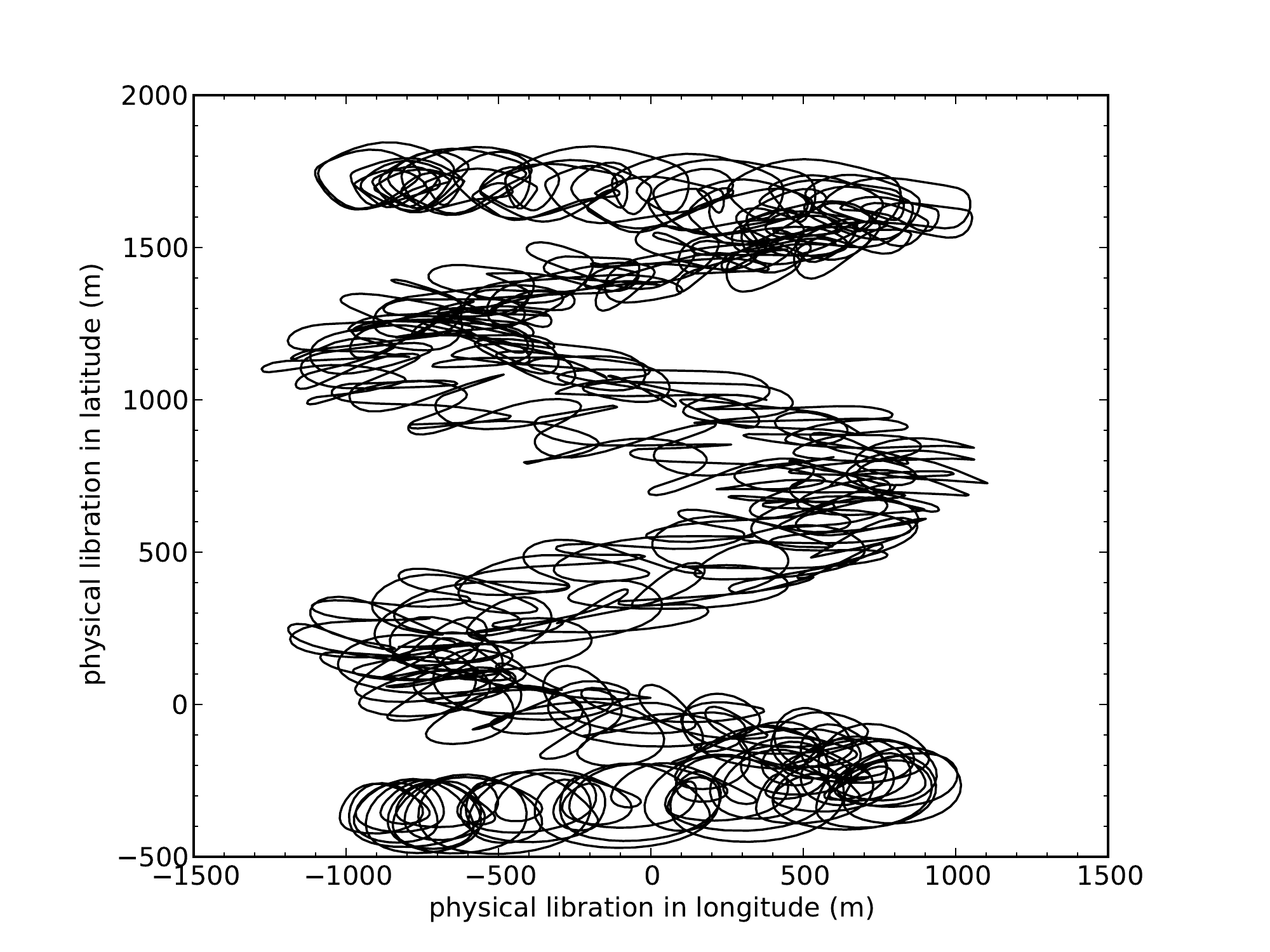}\end{center}

\caption{{\small Physical librations of the Moon over 18.6 years, translated into
displacement at the lunar surface. The main oscillation in longitude
has a one-year period, while the latitude exhibits a six year beat
period between the lunar anomalistic month and the draconic
month.\label{fig:physical-librations}}}
\end{figure}

The Moon is tidally locked to the Earth, with the equatorial bulge
elongated in the Earth-Moon direction, resulting in a triaxial mass
distribution---the associated moments of inertia labeled and ordered
as $A<B<C$. Torques on this non-spherical body from the Earth, Sun,
and larger/closer planets impose physical librations, or rocking,
of approximately $\pm120$~arcsec in both longitude and latitude.
This is distinct from the much larger ``optical'' libration caused
by Earth's changing vantage point of the Moon in its elliptical, inclined
orbit. At the surface of the Moon, the physical librations translate
into $\sim1000$~m amplitude motions (Fig.~\ref{fig:physical-librations}),
allowing centimeter-level LLR to gauge the effect at the $\sim10^{-5}$
level. Sensitivity to lunar physical librations has enabled determination
of relative differences in the principal moments of inertia, $\beta\equiv(C-A)/B$
and $\gamma\equiv(B-A)/C$ ($\sim6.3\times10^{-4}$ and $\sim2.3\times10^{-4}$,
respectively), to the 0.05\% level. In addition, the lunar quadrupole
moment $J_{2}$ couples to the lunar orbit and is fit from LLR data.
These three quantities together provide the best set of independent
measurements by which to determine the principal moments of inertia.
For instance, the polar moment, $C$, is found to be $0.393\pm0.001$
times $MR^{2}$, where a uniform density sphere would exhibit a numerical
factor of exactly 0.4 \cite{jgw-96-moon}. The Earth, by contrast,
is more centrally concentrated with a moment of inertia factor of
0.33.

If the Moon were a perfect fluid body, the tidal bulge caused by the
Earth would have a peak-to-trough amplitude of 19~m. However, the
Love number\footnote{Love numbers describe the degree to which a body
deforms relative to that of a perfect fluid body.  Rigid bodies would have
Love numbers near zero, while fluid bodies have Love numbers near unity.
The primary Love numbers, $h$, $l$, and $k$ describe vertical displacement,
horizontal displacement, and the degree to which the potential is modified
by the redistribution of mass, respectively.}, $h_{2}$, for the Moon is very small, at about 0.04 (contrast
to Earth $h_{2}\approx0.59$), so the total amplitude is held to less
than a meter. Optical librations limit the motion of the bulge relative
to the lunar surface to $\pm8^{\circ}$, so that any given spot on
the Moon sees tidal variations only at the level of $\sim0.1$~m,
peaking at 45$^{\circ}$ from the Earth-Moon line. Varying distance
between Earth and Moon contributes an additional $\pm0.1$~m deformation,
aligned with the bulge \cite{jgw-96-moon}. 

The most sophisticated and successful lunar interior model resides
at the Jet Propulsion Laboratory, for which the remainder of this
paragraph applies. In addition to $J_{2}$, $\beta$, and $\gamma$,
any combination of third-order multipole coefficients may be fit from
LLR data. Higher order terms and third-order terms that are not fit
derive from Lunar Prospector data. In the numerical integration of
lunar rotation, degree-2 Love numbers are considered, plus a time
delay for lunar tides. Dissipation is represented by a term for friction
at the core-mantle boundary and five out-of-phase periodic libration
terms address the frequency-dependence of tidal dissipation. At the
monthly tidal period, the dissipation is found to result in a rather
low resonant quality factor, $Q\approx33\pm4$. For extensive details
on the dissipation model, and treatment of physical librations, see
Williams et al. (2001) \cite{liquid-core}, and Rambaux and Williams
(2011) \cite{rambaux}, respectively.

\subsubsection{Earth Orientation, Precession, and Coordinate Frames\label{sub:Earth-Orientation,-Precession,}}

\begin{figure}[tbh]
\begin{center}\includegraphics[%
  width=4in,
  keepaspectratio]{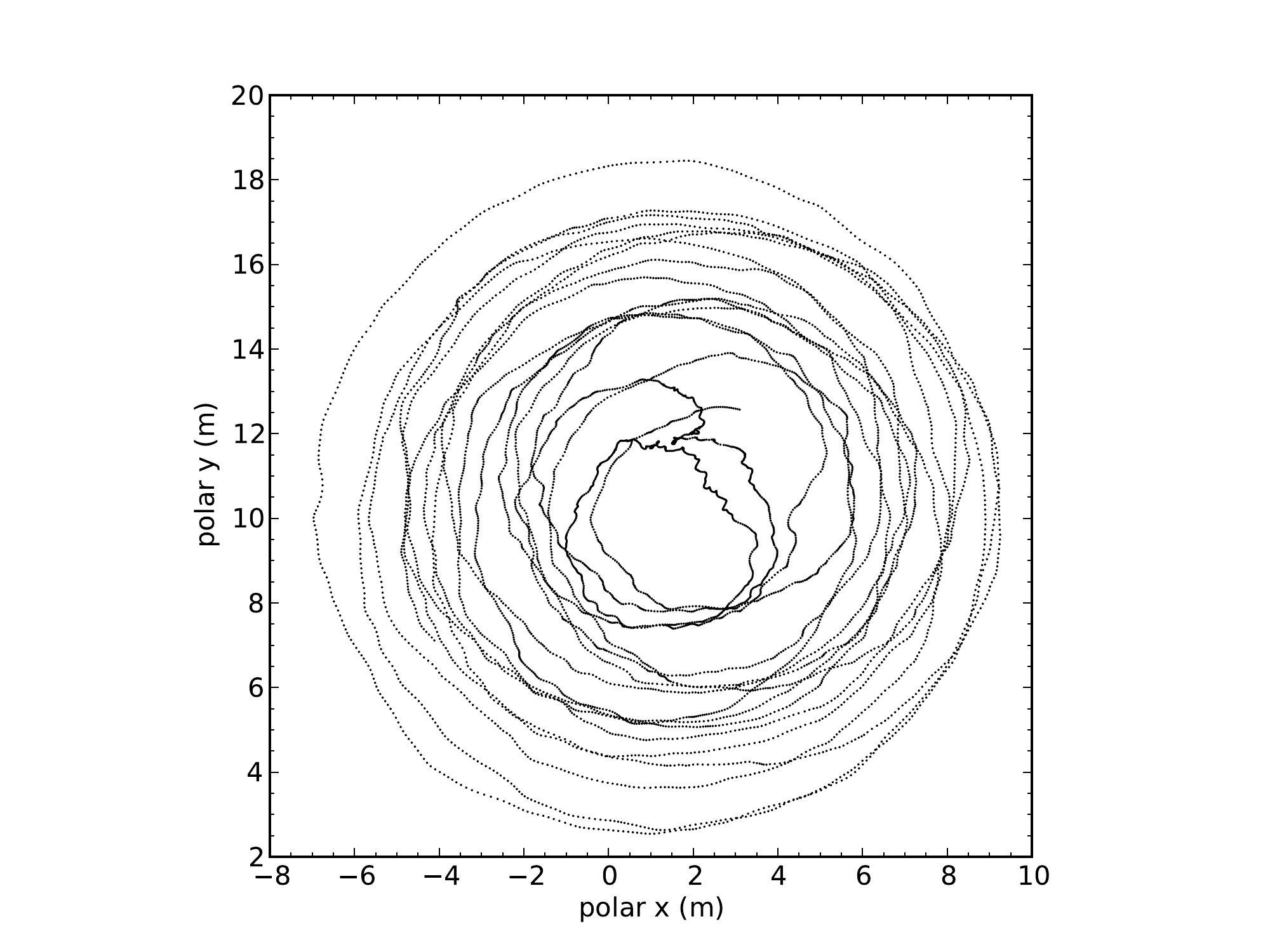}\end{center}

\caption{{\small Migration of Earth's polar axis with respect to the body over 20
years, in one-day samples.\label{fig:x-y-polar}}}
\end{figure}

Evaluating the center-to-center Earth-Moon range for access to
gravitational physics relies on detailed knowledge of the three-dimensional
orientation of the Earth at the time of observation. LLR can therefore
contribute to our knowledge and understanding of the responsible
phenomenologies.  Earth orientation can be described as three Euler angles,
broken into rotation about the polar axis, and two angles representing the
direction of said axis on the sky. Additionally, the rotation axis migrates
with respect to the geometrical surface on a 10~m scale over approximately
annual periods (this includes the Chandler
wobble\footnote{The Chandler
wobble is a free-mode (torque-free) nutation of the Earth's rotation axis
due to Earth's non-axisymmetric mass distribution.}; see
Fig.~\ref{fig:x-y-polar}).  The trajectory of the axis on the sky is
largely deterministic, described by precession and
nutation\footnote{Precession is the 26,000 year period migration of the
Earth's polar axis with respect to inertial space, sweeping out a cone
perpendicular to the ecliptic plane defined by Earth's orbit about the Sun.
Nutation is the much faster wobble of the polar axis about the smooth
precession trajectory due primarily to lunar and solar torques on the
Earth's equatorial bulge.} and related to known torques acting on the
Earth's figure. Nutation is dominated by an 18.6 year term relating to
precession of the lunar orbital plane, and amounting to the equivalent of
300~m at the Earth's surface. Earth's slowing rotation due to tidal
dissipation (accompanied by an LLR-determined 3.8~cm~yr$^{-1}$ egress of
the lunar orbit \cite{jgw-poznan}) results in a secular phase offset in the
rotation angle, also influenced by periodic phenomena like tides, and by
aperiodic factors like angular momentum exchange between ocean, atmosphere,
and land.

\begin{figure}[tbh]
\begin{center}\includegraphics[%
  width=4in,
  keepaspectratio]{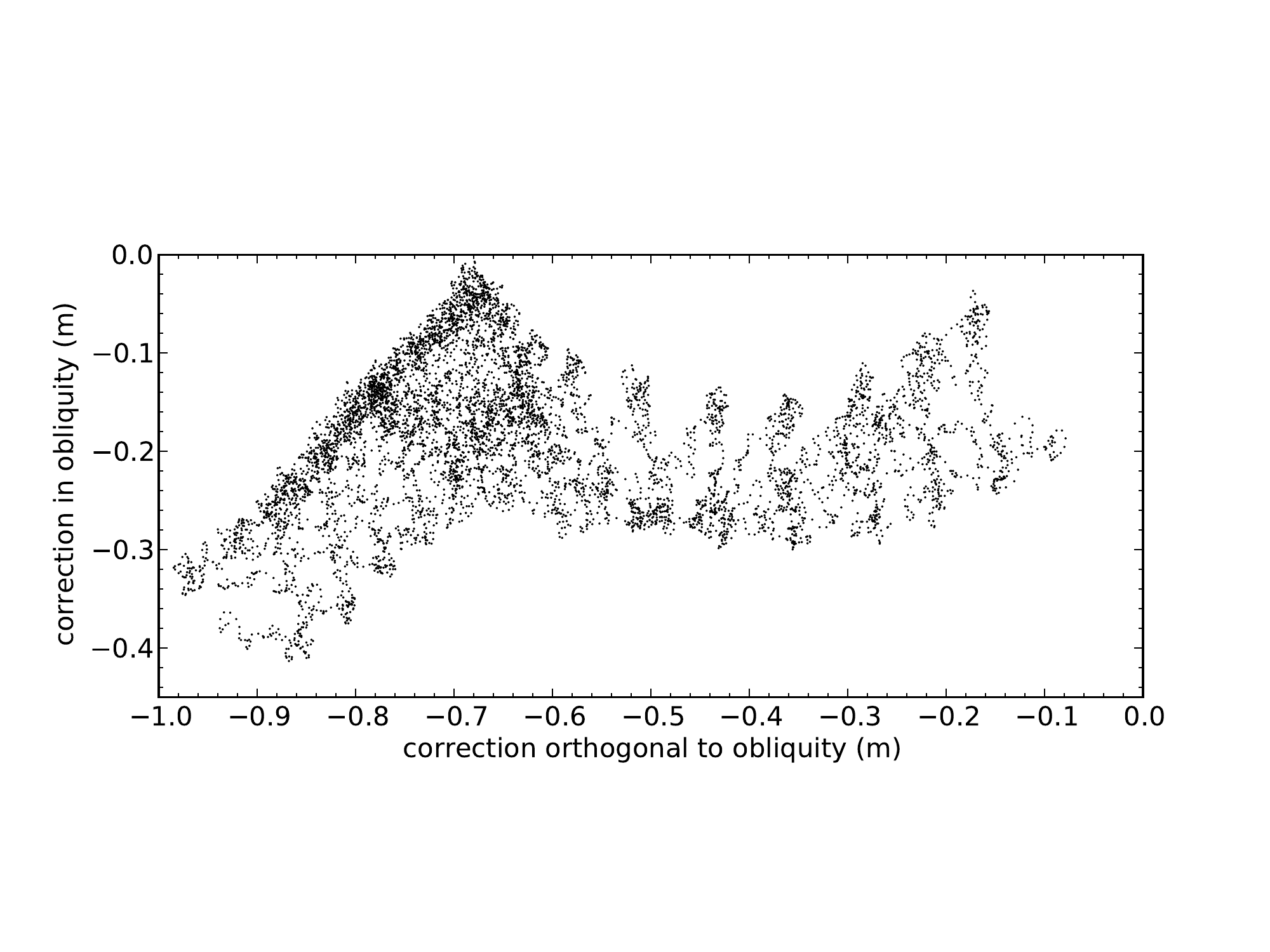}\end{center}

\caption{{\small Empirical nutation corrections over 20 years, relative to the
IAU1980 nutation theory, expressed in meters at the pole.
\label{fig:nutation-correct}}}
\end{figure}

\begin{figure}[tbh]
\begin{center}\includegraphics[%
  width=4in,
  keepaspectratio]{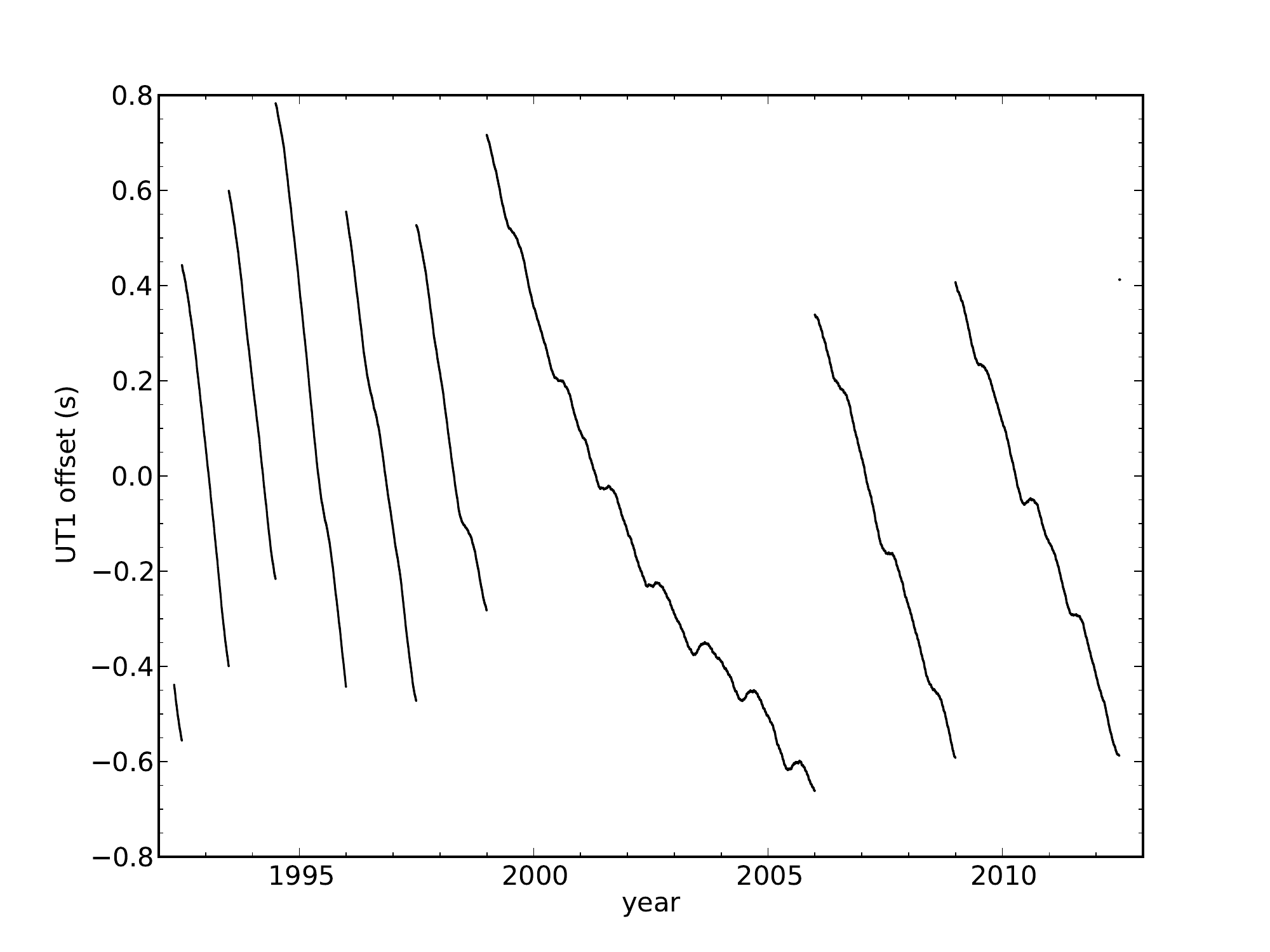}\end{center}

\caption{{\small UT1, the offset from Coordinated Universal Time (UTC), over 20 years.
Discontinuities reflect leap seconds. Annual wiggles can be seen in
all tracks, but most clearly when the slope is smaller, as in the
period from 2000--2005.\label{fig:UT1}}}
\end{figure}

\begin{figure}[tbh]
\begin{center}\includegraphics[%
  width=4in,
  keepaspectratio]{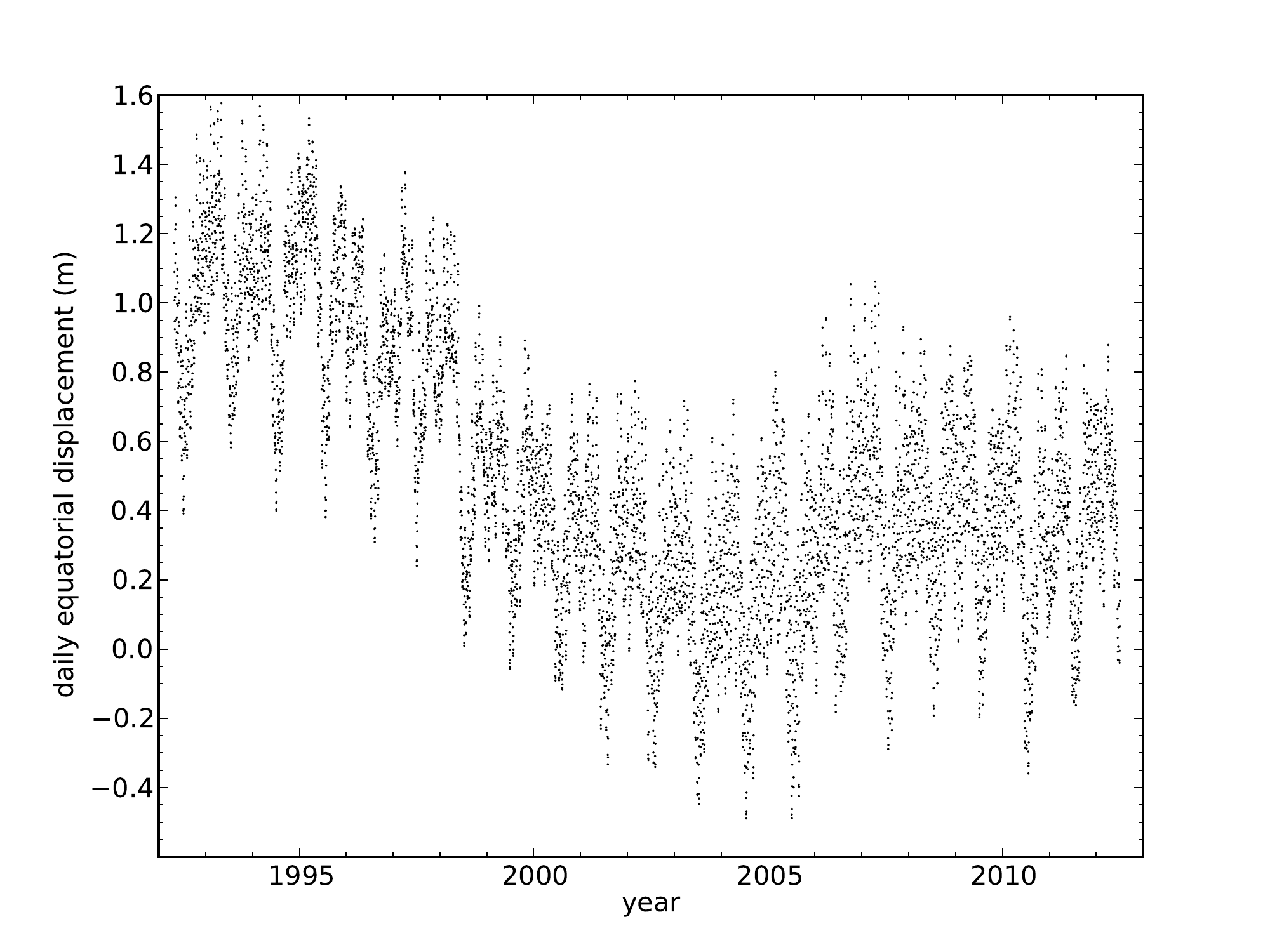}\end{center}

\caption{{\small Length of day variations expressed as the displacement of the equator
from one day to the next relative to a perfect 24-hour clock. The
net positive bias is simply a consequence of Earth's slowing rotation
due to tidal dissipation and is related to the 3.8~cm~yr$^{-1}$
egress of the lunar orbit. Large variations exist on top of this bias
at a level that is important to LLR.\label{fig:LOD}}}
\end{figure}

In practice, all five orientation parameters must be supplied or augmented
by observational measurements or corrections. Nutation adjustments tend to
be less than 1~m (Fig.~\ref{fig:nutation-correct}). UT1, characterizing the
rotational state relative to atomic time, can vary by as much as 3.5~ms in
a day (accumulating to almost one second per year around 1995), translating
to 1.6~m at Earth's surface (Figs.~\ref{fig:UT1} and \ref{fig:LOD}). The
point, again, is that LLR is sensitive to each of the Earth orientation
parameters, so that the LLR dataset can be used to supplement our
understanding of these phenomena (e.g., Biskupek and M\" uller 2009
\cite{biskupek}). LLR data are therefore routinely combined with data from
Very Long Baseline Interferometry (VLBI), GPS, satellite laser ranging
(SLR), and Doppler Orbitography and Radiopositioning Integrated by
Satellite (DORIS) in the Earth orientation parameters published by the
International Earth Rotation and Reference System Service (IERS), who also
periodically publish self-consistent, evolving methods for computing Earth
orientation \cite{iers-conv}.

Finally, LLR also contributes to establishment of coordinate systems,
especially in defining the relative orientation of the ecliptic and
equatorial planes\footnote{The ecliptic plane is the plane of the
Earth's orbit about the Sun.  The equatorial plane is a projection of the
Earth's equator into inertial space.}.  This in turn plays
a role in establishing the celestial coordinate origin, whether in the
International Celestial Reference System (ICRF) or the dynamical
ecliptic/equator of J2000.0 (whose coordinate origins differ by 17~mas).
Precession and nutation are defined with respect to the celestial frame, so
that LLR's sensitivity to the Earth axis orientation ties into this context
as well.

\section{LLR Capability across Time\label{sec:LLR-capability}}

Beginning with a historical introduction, this section looks at past and
current LLR capabilities, and how the current state of the art facilitates
improved scientific return from LLR.

\subsection{Brief LLR History\label{sub:history}}

Less than two weeks after the landing of the first retroreflector
array on the Apollo 11 mission, the first accurate laser ranges to
the Moon were performed on 1969 August 1 from the 3.1~m telescope
at the Lick Observatory. A few other sites around the world demonstrated
lunar ranging capability around that time, but none of these stations---including
the Lick Observatory---embarked on scientific campaigns to obtain
meaningfully long time series of accurate ranges. However, one month
after the Apollo~11 landing, a long-term effort using the 2.7~meter
telescope at the McDonald Observatory commenced ranging to the Moon
\cite{bender}, providing all of the scientifically relevant observations
over the next decade. The McDonald station used a ruby laser with
a 4~ns pulse width, firing at a repetition rate of about 0.3~Hz
and $\sim3$~J per pulse. This station routinely achieved 20~cm
range precision, with a photon return rate as high as 0.2 photons
per pulse, or 0.06 photons per second. A typical ``normal point''---a
representative measurement for a run typically lasting tens of minutes---was
constructed from approximately 20 photon returns.

In the mid 1980's, a lunar ranging renaissance took place, with three
capable stations beginning operation. In 1984, a French station at
the Observatoire de la C\^ote d'Azur (OCA) \cite{oca} began collecting
accurate ranges. Using a 1.5~meter telescope, a 70~ps Nd:YAG laser
firing at 10~Hz and 75~mJ per pulse, OCA became the premier lunar
ranging station in the world and has contributed about half of the
total range measurements to date. From 1984--1990, a station at Haleakala
in Hawaii produced strong returns and accurate ranges. In 1985, the
McDonald operation transitioned from the 2.7~m shared astronomical
telescope to a dedicated 0.76~m telescope (also used for satellite
laser ranging) using a 200~ps Nd:YAG laser operating at 10~Hz and
150~mJ per pulse. This station is referred to as the McDonald Laser
Ranging System (MLRS) \cite{mlrs}. From 1990--2006, the MLRS and
OCA stations were the only routine contributors to lunar range data
with characteristic return rates of 0.002 and 0.01 photons per pulse,
respectively. Normal points from the two stations typically consist
of 15 and 40 photons, respectively. More complete histories of these
and other efforts (Russia, Pic du Midi, Australia, Japan, etc.) may
be found in other works \cite{jgw-ep,dickey,bender}.

Presently, five stations in the world exhibit LLR capability: OCA,
MLRS, Apache Point, Matera, and Wetzell---although only the first
three acquire data regularly. After a shutdown from 2005--2010, OCA
is back in regular operation, although at about half its former pace.
Since 2007, the Apache Point Observatory Lunar Laser-ranging Operation
(APOLLO; for which the author is the principal investigator) has led
the LLR data effort both in terms of number of normal points and estimated
range uncertainty. Averaging about 260 measurements per year and a
median statistical uncertainty per normal point of less than 3~mm,
APOLLO seeks to effect a substantial improvement in LLR tests of gravity.

\subsection{APOLLO Apparatus and Performance\label{sub:APOLLO-Apparatus}}

This section provides a brief overview of the APOLLO apparatus and its
demonstrated performance.  A full description of the apparatus can be found
in Murphy et al. (2008) \cite{apollo}.  APOLLO employs a laser averaging
2.3~W at 532~nm, generating 100~ps pulses at a 20~Hz repetition rate and
115~mJ per pulse. The laser is transmitted from the 3.5~m aperture
telescope at the Apache Point Observatory in southern New Mexico at an
elevation of 2.8~km. The full aperture is utilized for beam transmission. A
small portion of the outgoing beam is intercepted by a corner cube prism
attached to the telescope secondary mirror, sending light back to the
receiver, attenuated to the single-photon level and providing a precise
measure of the pulse departure time.  The receiver houses a $4\times4$
avalanche photodiode (APD) array capable of high-precision timing of single
photons at a detection sensitivity around 30\%. The array occupies a
re-imaged focal plane of the telescope, spanning 1.4~arcsec on a side. This
arrangement results in an oversampled point spread function, while
providing spatial information useful for tracking feedback. Photon arrivals
create START pulses for a 16-channel time-to-digital converter (TDC) with
15~ps jitter and 25~ps bins. STOP pulses to the TDC are extracted from a
50~MHz low-phase-noise clock pulse train, and the number of clock pulses
between the STOP signal for the local corner cube return and the STOP
signal for the lunar return is counted. The master clock on which the
50~MHz pulse train is generated uses an ovenized quartz crystal disciplined
by reference to GPS so that the 2.5~s round-trip travel time is measured
against a reliable frequency standard, and the absolute time is known far
better than the microsecond level required for millimeter range precision.
Table~\ref{tab:error-budget} reproduces the contributions to APOLLO's
random uncertainty from the instrument description paper \cite{apollo}.

\begin{table}
\caption{APOLLO Random Error Budget per Photon.\label{tab:error-budget}}
\begin{center}
\begin{tabular}{lcc}
\br
Error Source & RMS Error (ps) & RMS Error (mm) \\
\mr
APD illumination         & 60       & 9 \\
APD intrinsic            & $<50$    & $<7.5$ \\
Laser pulse              & 45       & 7 \\
Timing electronics       & 20       & 3 \\
GPS clock                & 7        & 1 \\
\mr
Total APOLLO             & 93       & 14 \\
Retroreflector array     & 100--300 & 15--45 \\
\mr
Total random uncertainty & 136--314 & 20--47 \\
\br
\end{tabular}
\end{center}
\end{table}

The large telescope aperture, good atmospheric seeing, and array detector
together result in high signal rates and allow confident signal
optimization through pointing corrections and velocity aberration
compensation---controlled by affecting a deliberate offset in transmitter
and receiver pointing directions. Consequently, APOLLO signal rates exceed
those of previous stations by a substantial margin. Table~\ref{tab:records}
displays APOLLO's record performance on each reflector.  Since each
entry is associated with 5000-shot data runs, each transpired over
approximately 250 seconds, at a 20 Hz pulse repetition rate.  The
``Photons" column corresponds to detected photo-electrons, sometimes
exceeding one photon per shot, as enabled by APOLLO's multi-element
APD detector.  The ``rate factor'' compares APOLLO's peak photon rate
(photons/minute) to that
of the previous LLR record for each reflector, held in every case by 
OCA---except for Lunokhod~1, which was first recovered by APOLLO, as described
in Section~\ref{sub:Finding-Lunokhod-1}.

\begin{table}

\caption{APOLLO Record Rates.\label{tab:records}}

\begin{center}\begin{tabular}{cccccc}
\br 
Reflector&
Shots&
Photons&
photons/shot&
photons/minute&
rate factor\\
\mr
Apollo 11&
5000&
4784&
0.96&
1148&
69\\
Apollo 14&
5000&
7606&
1.52&
1825&
69\\
Apollo 15&
5000&
15730&
3.15&
3775&
67\\
Lunokhod 1&
5000&
2070&
0.41&
497&
---\\
Lunokhod 2&
5000&
1301&
0.26&
312&
54\\
\br
\end{tabular}\end{center}
\end{table}

Greater photon count is not in itself indicative of higher precision
range measurements. But to the extent that the temporal spread of
the lunar return is dominated by the finite size of the reflector
array---tilted by lunar libration---high photon number is a necessary
ingredient in reducing statistical uncertainty, as was illustrated
in Section~\ref{sub:Reflectors-and-Divergence}. Two different evaluations
of APOLLO's millimeter-level performance have been published elsewhere
\cite{battat-mm,apollo-cqg}.

\subsection{APOLLO Advantages Translated to Science\label{sub:APOLLO-advantages}}

The aforementioned signal optimization capability of APOLLO naturally
translates into faster acquisition and a higher signal rate (see
Fig.~\ref{fig:A15-run} for an example), which enhance the range precision
and scientific usefulness of the data. But a number of derivative
advantages emerge as well, outlined here. As a general statement,
systematic effects are more easily exposed in a high-signal regime.

\begin{figure}[tbh]
\begin{center}\includegraphics[%
  width=6in,
  keepaspectratio]{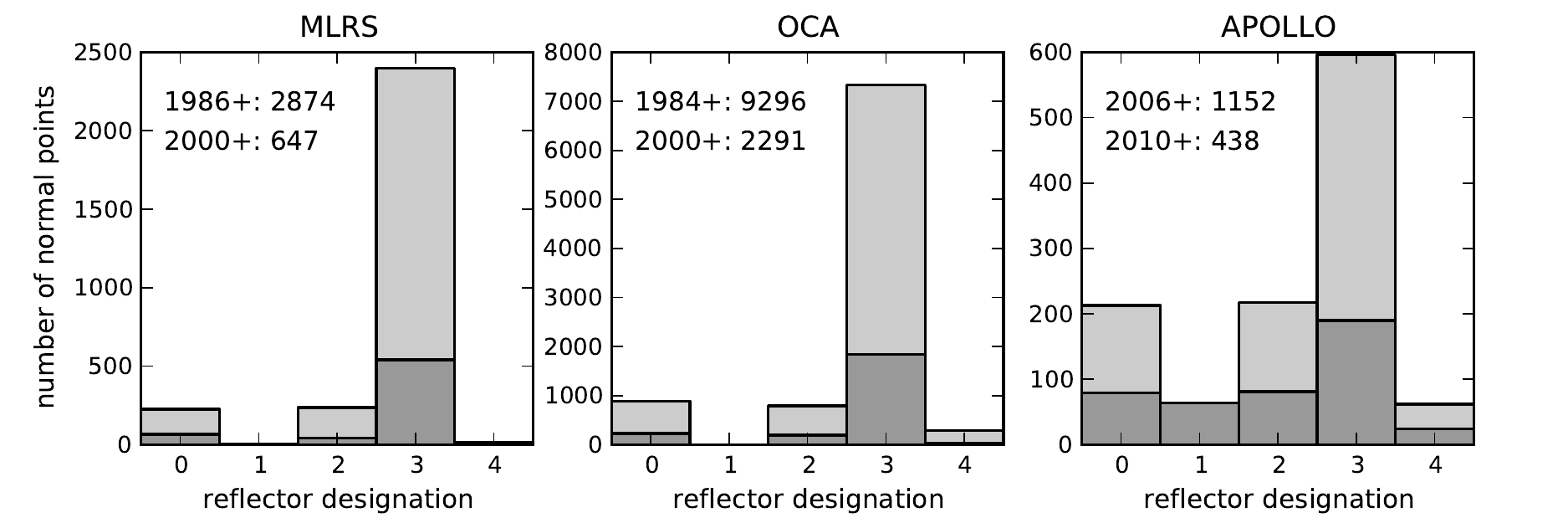}\end{center}

\caption{{\small Distribution of normal points among the five reflectors for the chief
modern LLR stations. Reflectors are numbered in order of arrival from
0--4, corresponding to Apollo~11, Lunokhod~1, Apollo~14, Apollo~15,
and Lunokhod~2. Overlaid on each is a darker histogram representing
more recent data, the break point being 2000.0 for the longer-lived
stations and 2010.0 for the newer APOLLO. Apollo~15 dominates for
all stations, being a larger, more easily acquired
target.\label{fig:np-refl}}}
\end{figure}

\begin{figure}[tbh]
\begin{center}\includegraphics[%
  width=6in,
  keepaspectratio]{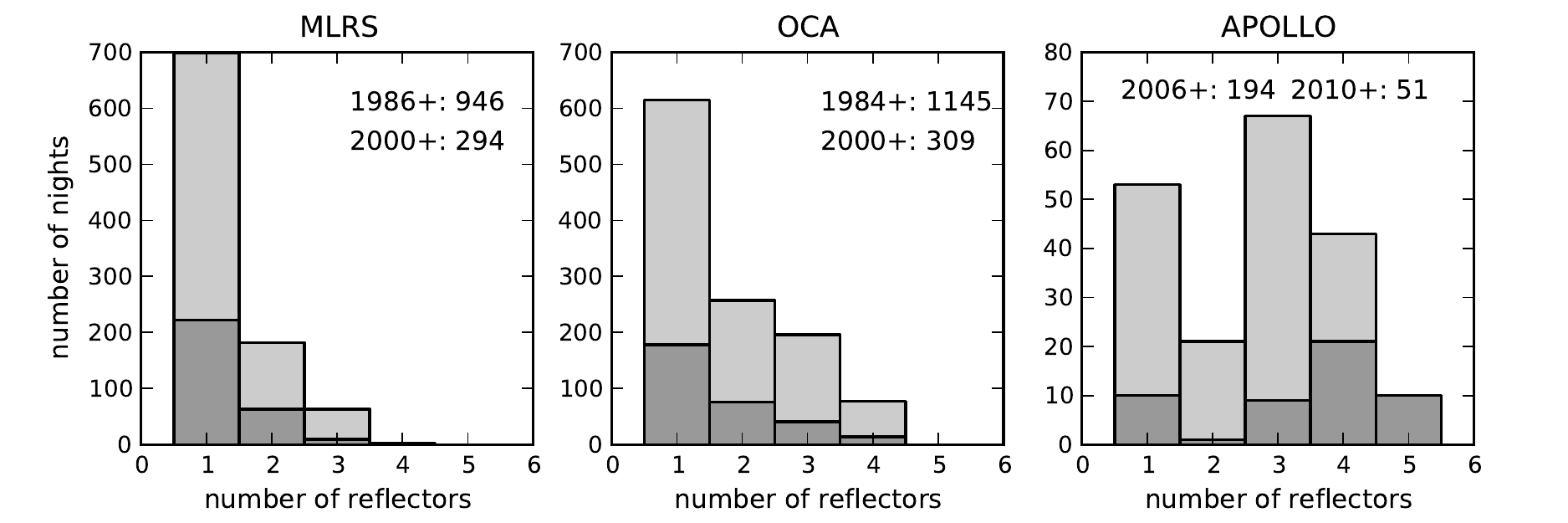}\end{center}

\caption{{\small Number of reflectors acquired per night of observation for the chief
modern LLR stations. Shading follows the convention of Fig.~\ref{fig:np-refl}
The mode for the older stations has been one reflector per night (both
overall, and recently), whereas APOLLO's nominal mode has been three
reflectors, recently moving to four.\label{fig:refl-spread}}}
\end{figure}

Foremost, APOLLO routinely ranges to four, and sometimes five reflectors
in each observing session. Typically, it is possible to make several
circuits of the reflectors within the $\sim1$~hr time allocation
(Figs.~\ref{fig:np-refl} and \ref{fig:refl-spread}). The result
is that lunar orientation and deformation are well-established during
each session. This bestows an obvious advantage onto understanding
of the lunar interior, but also enhances the ability to represent
accurately the location and trajectory of the center of mass of the
Moon---important for testing gravitation.

Having multiple detector channels essentially provides many independent
measurements: each channel receives photons from the local (fiducial)
corner cube as well as from the lunar array. One may then compare
``answers'' from each of the channels to get a separate handle
on measurement error. The degree to which measurements disagree provides
a check on estimated uncertainties.

The signal rate is high enough to permit exploration of the physical
orientation of the reflector arrays on the Moon via the evolution
of the temporal response as a function of lunar libration. This is
especially effective on the larger, rectangular Apollo~15 array,
which appears to have a 2$^{\circ}$ azimuth offset, but otherwise
nominal tilts. The smaller Apollo arrays are consistent with nominal
pointing, although probed less accurately. Incorporating array orientation
information into the data reduction routines allows a higher fidelity
fit to the observations, reducing the opportunity for systematic offsets.
Ultimately, it may be possible to elucidate spatial variability of
response across the reflector array.

\section{Recent Surprises\label{sec:Recent-Surprises}}

Recent improvement in the LLR return rate has facilitated some new
findings, two of which are detailed here: first the re-discovery of a
reflector that had been lost for nearly four decades; followed by an
account of evidence for degraded reflector performance.

\subsection{Finding Lunokhod 1\label{sub:Finding-Lunokhod-1}}

Starting in 2008, APOLLO devoted some observing time to searching
for the lost Lunokhod~1 reflector. Accurate coordinates were not
available, so searches were referenced to the best-guess coordinates
of the time \cite{stooke}. Given a $\sim2$~km beam footprint, the
5~km positional uncertainty can be scanned in a matter of minutes,
so that angular parameter space was not the main limitation. Rather,
the 100~ns timing gate used for the APD detector array translates
into a $\pm7.5$~m line-of-sight uncertainty, intersecting the tilted
lunar surface at the position of Lunokhod~1 in a $\sim20$~m swath.
Therefore searches concentrated exclusively on the temporal domain
while pointing at the nominal position and allowing natural pointing
excursions to provide some degree of angular coverage.

\begin{figure}[tbh]
\begin{center}\includegraphics[%
  height=2.5in,
  keepaspectratio]{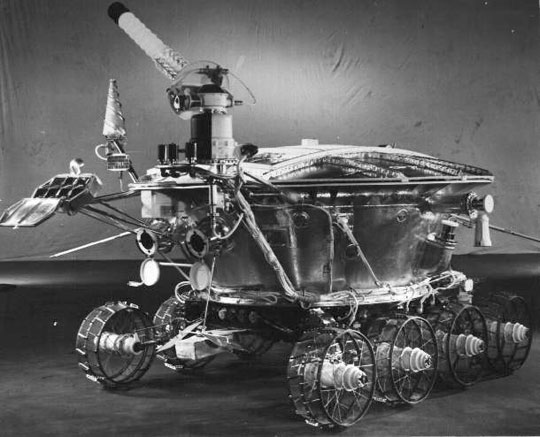}\hfill
  \includegraphics[%
  height=2.5in,
  keepaspectratio]{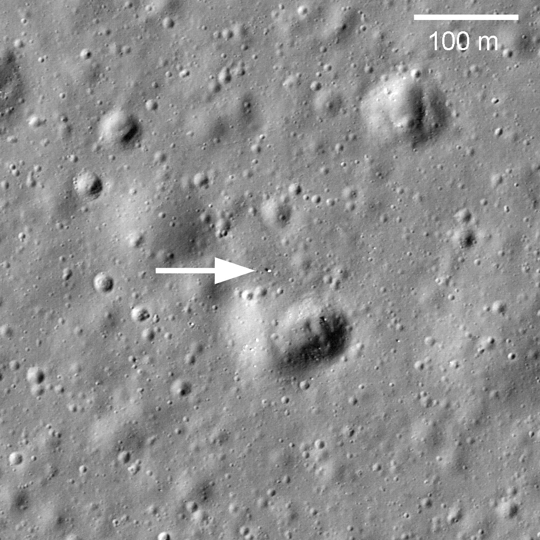}
\end{center}

\caption{{\small Two Lunokhod rovers, one of which is pictured at left,
were landed and operated on the lunar surface, both possessing a
retroreflector array (visible as a tilted tray protruding from the rover body at far
left; image courtesy the Lavochkin Association).  Any records of successful ranging to the first rover were lost to
the international community, so that accurate coordinates were unknown, and
nearly 40 years passed without range measurements.  In 2010 March, NASA's
Lunar Reconnaissance Orbiter Camera (LROC) obtained the image at right,
locating the rover (arrowed) to approximately 100~m accuracy.  Successful
ranging has thenceforth been possible, the reflector appearing to be in
better health than its twin on the Lunokhod~2 rover.  \label{fig:Lunokhod}}}
\end{figure}

The Lunar Reconnaissance Orbiter (LRO) helped in three ways. Most
indirectly, corner cube prisms were placed on LRO in the hope that
APOLLO could obtain 2-way range measurements to the spacecraft. This
prompted APOLLO to develop a wide-gate (800~ns) mode to cope with
positional uncertainty of LRO. Second, LRO's Lunar Orbiter Laser Altimeter
(LOLA) provided an accurate radial coordinate of the plain on which
Lunokhod~1 sits. But by far the most important contribution was high-resolution
imaging, identifying the rover in 2010 March, and providing coordinates
accurate to approximately 100~m (Fig.~\ref{fig:Lunokhod}).

On 2010 April 22, APOLLO got a strong return from the Lunokhod~1
reflector, appearing 270~ns later than the time prediction based
on the LRO-provided coordinates. Recording about 2000 photons in the
initial 10,000 shot run, Lunokhod~1 had instantly outperformed the
best photon yield APOLLO had seen over five years of observing the
Lunokhod~2 reflector (in 44 prior measurements). Since this time,
the position---off from the earlier working estimate by 5~km, and
off from the LRO-provided coordinates by 100~m---has been refined
to the centimeter level \cite{lunokhod}.

The Lunokhod~1 reflector is located $50^{\circ}$ from the selenographic
coordinate origin, making it the farthest reflector from the apparent
lunar center: about twice as far as the Apollo reflectors (Fig.~\ref{fig:locations}).
This makes Lunokhod~1 a more sensitive probe of lunar orientation
than the other reflectors. Moreover, its location allows sensitivity
to librations in both latitude and longitude, while the Apollo reflectors
lie close to the equator or prime meridian, resulting in each being
mostly sensitive to librations in only one principal dimension.

\subsection{Reflector Degradation\label{sub:Reflector-Degradation}}

Despite confident optimization of telescope pointing, velocity aberration
correction, laser beam divergence, and telescope focus (see
Section~\ref{sub:APOLLO-Apparatus}), the APOLLO
link budget has never matched its theoretical potential---falling
short by an order of magnitude, even on the best nights \cite{canberra-thruput}.
This appears to be true for other LLR stations as well, based on comparing
performance to expectations with respect to system parameters. More
telling is the observation that the signal level near full moon phase
drops by another order of magnitude \cite{dust}. Examination of the
earliest range data from the McDonald 2.7~m telescope reveals the
slow onset of this phenomenon, so that the cause appears to be progressive
in nature.

Lunar eclipse observations contribute a substantial clue, in that
the signal performance soars to normal levels within 15 minutes of
entering full shadow. This strongly suggests solar energy absorption
leading to thermal gradients in the corner cube prisms. The Apollo
corner cubes and associated mounts were carefully designed to minimize
solar absorption and thermal gradients by a combination of total internal
reflection, recessed corner cubes, radiation-resistant substrate,
and minimization of mount conductance. Thermal simulations predicted
a central irradiance of the far-field diffraction pattern emerging
from the corner cubes at full moon to be $>90$\% of the peak performance.
A gradient as small as $\sim4$~K from the front surface to the corner
cube vertex essentially nullifies the central irradiance \cite{gradients}.

The most likely explanation is the slow accumulation of a very thin
layer of dust on the reflector front surface, transported by electrostatic
levitation (via photoionization and solar wind charge deposition \cite{stubbs-dust,farrel-dust})
and cascading disturbances from impacts. Approximately half of the
reflector surface would need to be covered by dust to produce a ten-fold
reduction in central irradiance, since each dust grain counts twice
in a double-pass of the front surface, and the central intensity otherwise
scales as the square of the clear area. Meanwhile, this same $\sim50$\%
fill-factor could result in enough front-surface thermal absorption
to generate a thermal gradient sufficiently large to cause an additional
large signal deficit.

Indeed, the eclipse observations validate this picture, in that after
the initial signal surge upon entering shadow, the return strength
plummets to sub-detectable levels. When light returns, the signal
peaks again before settling back to levels typical for the full moon.
The interpretation is that initially the corner cube has a strong
positive thermal gradient owing to a heated front surface. As the
solar illumination fades, the corner cube begins to radiate its stored
energy to space via the front surface, cooling off and reversing the
gradient. A zero-crossing occurs as the gradient evolves from positive
to negative, so that the reflector performance momentarily recovers
during the approximately isothermal state. When light returns, the
gradient changes sign again, passing through zero a second time for
a temporary surge in signal strength.

One final aspect of the degradation story is that Lunokhod~2 was
initially observed in 1971 to be comparable in strength to the Apollo~15
reflector. Indeed, cross section predictions put the expected Lunokhod
response midway between the 100-element Apollo~11/14 reflectors and
the 300-element Apollo~15 reflector. Today, Lunokhod~2 registers
at about 10\% the strength of the Apollo~15 array. Thus it is clear
that the reflectors can experience \emph{relative} changes in performance
over time. Meanwhile, the three Apollo reflectors are observed to
maintain a 3:1:1 ratio, and all exhibit a comparable full-moon deficit.
The Lunokhod~1 array typically performs similarly to---if not better
than---the smaller Apollo arrays. Its degradation therefore appears
to be roughly in step with that of the Apollo reflectors, making the
identically-designed Lunokhod~2 reflector the truly anomalous case.

The possibility of dynamic dust on the Moon impacts ambitions for
lunar-based telescopes or mechanical equipment. The still-functioning
reflectors have often been held up as evidence that dust is not a
major issue. The recent observations described here cast doubt on
this picture.

\section{The Modeling Challenge\label{sec:The-Modeling-Challenge}}

Sitting between accurate range measurements and scientific results
is a complex model whose development must keep pace with observational
advances in order to realize the full potential of LLR. Here we describe
the requisite components of a model, discuss current capabilities,
and explore improvements to be made.

\subsection{Model Content/Construction\label{sub:Model-content/construction}}

The crux of any scientific endeavor is the comparison between theory
and experiment. For LLR, the theory piece is represented by a model
of the solar system incorporating some prescription for gravity and
all other physical effects that can render an impact on the measurement.
For example, Venus and Jupiter generate perturbations in the Earth-Moon
separation on the order of 1~km, Mars and Saturn at the 100~m level,
and even the largest asteroids chalk up millimeter-scale deviations.
It is therefore clearly important to have an accurate representation
of solar system dynamics.

But because LLR is performed in relation to the \emph{surfaces} of
the Earth and Moon, it is also necessary to provide accurate descriptions
of body orientations and deformations. In the case of the Earth, non-deterministic
mass flows in the atmosphere and ocean complicate matters. Body torques
between Earth and Moon not only affect orientation, but also couple
into orbital dynamics. Tidal dissipation likewise translates into
an orbital egress of the Moon at the rate of about 38~mm~yr$^{-1}$.
Crustal loading influences from the ocean, atmosphere, and ground
water come into play for the Earth station.

Light propagation effects must also be considered. Sensibly cast in
the solar system barycenter frame (see Section~\ref{sub:Fundamental-Measurement-and}),
the light path forms two legs of a generally asymmetric triangle.
The Shapiro delay associated with propagation through solar and terrestrial
gravitational potentials must be incorporated, amounting to a $\sim25$~ns
modification to the round-trip time (7.5~m one-way equivalent due
to the Sun; 0.04~m from the Earth potential). Propagation through
the atmosphere incurs a roughly 2~m path delay that must be determined
and removed to high precision.

The model itself is constructed as a parameterized physical description,
many pieces of which are numerically integrated simultaneously. For
example, solar system bodies are represented as point masses, where
model parameters are initial positions and velocities, and the associated
mass values. The Earth and Moon and Sun are treated as non-point masses,
in which case the dynamical torques are jointly computed to follow
the dynamical evolution of the system. Partial derivatives of the
computed range with respect to each model parameter are calculated
for each measurement epoch so that a least-squares covariant parameter
adjustment may be executed. By iterating such adjustments, the set
of LLR observations can be used to converge on a physical description
of the solar system that is optimally consistent with the data. A
cute way to put this is that millimeter-level measurement/model fidelity
in the Earth-Moon range can in principle determine the mass and position
of Jupiter to a part in a million, given the kilometer-scale influence
Jupiter has on the lunar orbit. In practice, irregular data sampling
together with correlations between many model parameters compromise
complete separation of variables. Over the long term, periodic effects
from solar system bodies tend to be separable. But some parameters
tend to remain highly correlated, like the $GM$ value for the Earth-Moon
system and the semi-major axis of the lunar orbit.

\subsection{Current Capabilities\label{sub:Current-Capabilities}}

Several LLR models exist in the world, sited at the Jet Propulsion
Laboratory (JPL), the Harvard-Smithsonian Center for Astrophysics
(CfA), the Leibniz University in Hannover, Germany, and at the IMCCE
in Paris, France. Of these, only the Planetary Ephemeris Program (PEP),
at the CfA, is made available to the community (and as open-source
code). The JPL model currently demonstrates the best performance,
producing weighted RMS residuals for both APOLLO and OCA data in the
neighborhood of 18~mm, which is roughly a factor of two better than
the other models at present. Clearly a gap exists between estimated
APOLLO uncertainties of a few millimeters and the model residuals.

The CfA, Paris, and Hannover efforts are currently engaged in a stepwise
comparative effort to identify model differences, shortcomings, and
errors. Additionally, APOLLO data are being used to illuminate one
aspect of model performance by exploiting the fact that most observing
sessions result in measurements to multiple reflectors. This provides
a nearly direct measurement of lunar orientation, the stability of
which is confirmed on occasions when several circuits of the reflectors
are made in a short period \cite{apollo-cqg}. In brief, the result
of this exercise is a determination of how much adjustment is needed
in the latitude and longitude librations of the Moon to bring the
residuals among the reflectors in line with each other. The results
are summarized in Table~\ref{tab:lib-adjust}. For reference, 1~nrad
of angle translates to 1.7~mm of range at the lunar limb, or about
0.7~mm at the typical position of an Apollo reflector. Again it is
clear that JPL has an advantage over the other efforts, although the
Paris result does not represent a least-squares-adjusted integration,
but rather uses the JPL exported ephemeris, DE423.

\begin{table}

\caption{Libration adjustment weighted RMS in nanoradians.\label{tab:lib-adjust}}

\begin{center}\begin{tabular}{lcc}
\br
Model&
longitude adjustment (nrad)&
latitude adjustment (nrad)\\
\mr
JPL&
5.1&
7.3\\
CfA&
19&
23\\
Hannover&
23&
36\\
Paris&
29&
76\\
\br
\end{tabular}\end{center}
\end{table}

Lunar orientation is one of many components in the model, so that
adjusting the librations in an ad-hoc manner based on APOLLO residuals
does not markedly improve the overall RMS of residuals---the main
effect being to better cluster residuals from different reflectors
within each night. Night-to-night variations still dominate, and tend
to look wholly different from one model to the next.

\subsection{Charted Improvements\label{sub:Charted-Improvements}}

Each LLR analysis group has its own list of known effects yet to be
incorporated into the model---many of which are only beginning to
be important at the millimeter level. By way of example, the following
is a list of known effects not yet incorporated into PEP at CfA. Other
groups may be in different states with regard to these items. For PEP, specific
improvements to be made include:

\begin{itemize}
\item A more complete treatment of dissipation in the lunar interior, following
JPL's lead;
\item A more rigorous tidal model, applying Love numbers that depend on
frequency and spherical harmonic degree and order, aided by inputs
from VLBI and GPS;
\item Updating the gravitational multipoles of the Earth and Moon, using
the latest data from the GRACE and GRAIL missions, respectively;
\item Improved Earth orientation handling, including feedback of LLR residuals
into the VLBI/GPS-determined data;
\item Ocean loading, having approximately 3~mm horizontal RMS and 5~mm
vertical RMS at the APOLLO site, for instance;
\item Atmospheric loading, having an impact of roughly 1~mm for every 3~mbar
of pressure anomaly;
\item Earth center-of-mass motion, seen via SLR
to have a $\sim1$~cm amplitude at an annual frequency.
\end{itemize}
Addressing the first three items---together with any errors discovered
in the model intercomparison between PEP and the European models---may
in fact bring PEP in line with the present JPL model capability. The
remaining effects could conceivably add up to accommodate the $\sim2$~cm
residuals still exhibited by the JPL model. We discuss them here.

Sophisticated models exist for tidal ocean loading that describe site
motion at the sub-millimeter level---easily so for APOLLO, since the
semi-diurnal load tides at Apache Point happen to be small. Table~\ref{tab:ocean}
presents modeled site displacements from ocean tidal loading at the
Apache Point site by the TPXO 7.0 model. Other models (GOT00 and CSR4)
produce results consistent to within about 0.5~mm.

\begin{table}[tbh]

\caption{Ocean Loading Amplitudes at Apache Point.\label{tab:ocean}}

\begin{center}\begin{tabular}{lccc}
\br
Component&
RMS (mm)&
Minimum (mm)&
Maximum (mm)\\
\mr 
North-South&
$2.26$&
$-5.88$&
$5.57$\\
East-West&
$1.57$&
$-3.52$&
$4.51$\\
Vertical&
$5.46$&
$-14.71$&
$11.91$\\
\br
\end{tabular}\end{center}
\end{table}

For the various sources of non-tidal loading, the best results will
come from a combination of the global pressure fields produced by
the various branches of the Global Geophysical Fluids Center \cite{ggfc}:
air pressure, ocean mass, and ground and surface water. Global models
for these are advancing rapidly thanks to data from the GRACE mission
\cite{Tregoning05,Ramillien08}. These can be improved by combining,
for example, global models of air pressure with the more detailed
(``mesoscale'') local models that are now produced for regional weather
forecasting.

Satellite laser ranging measurements show a displacement of Earth's
center-of-mass with respect to coordinates of the geometrical center
(as defined by a network of ground stations). This motion has a roughly
annual period and an amplitude of about a centimeter \cite{cmmotion,SLR-COM,GPS-COM}.
LLR analysis has not yet incorporated this effect. However, the presence
of the nearby a GPS station (described in Section~\ref{sub:Uncharted-Improvements})
will allow us to incorporate the SLR result on geocenter motion into
the analysis of APOLLO data.

Additionally, radiation pressure is known to be a $3.65\pm0.08$~mm
$\cos D$ effect that can be applied \cite{radpress}. Likewise, atmospheric
propagation delay, a $\sim2$~m effect, has been recently modeled
to sub-millimeter accuracy for elevation angles above 20$^{\circ}$
\cite{mendes-zenith,mendes-mapping}, and has since been incorporated
into PEP.

\subsection{Uncharted Improvements\label{sub:Uncharted-Improvements}}

As described, the items in the previous section are moderately well
understood, in most cases having mature models and complementary
observations providing input. But millimeter-quality LLR data will likely
strain current models and demand that new physical effects be addressed.

\begin{figure}[tbh]
\begin{center}\includegraphics[%
  width=6in,
  keepaspectratio]{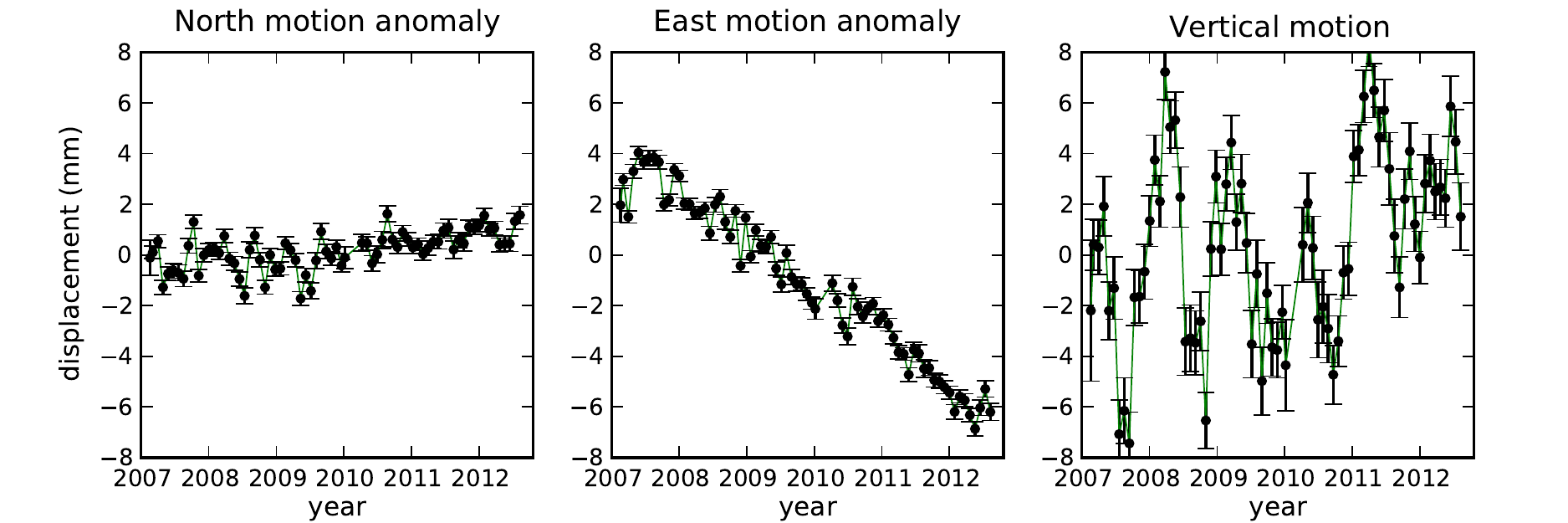}\end{center}

\caption{{\small GPS data from the P027 Plate Boundary Observatory station
located 2.5 km away from the Apache Point Observatory (on a similar
summit). Data are binned in lunar-monthly units. Motions are shown}
\emph{\small relative to} {\small the North American plate, moving
at ($-6.3$, $-11.5$, $-0.6$) mm/yr in the north, east, and up directions.
The net motion of station P027 with respect to the global frame becomes
($-6.05$, $-13.35$, $+0.2$) mm/yr. The vertical motion indicates
peak-to-peak site displacements exceeding 1~cm, highlighting the
need to incorporate geodetic measurements into millimeter-quality
LLR analysis.\label{fig:GPS}}}
\end{figure}

APOLLO is provisioned to check how well displacements are being modeled
by comparison to supplemental measurements of: 

\begin{itemize}
\item site position by a nearby GPS station (P027 in the Plate Boundary
Observatory network; 2.5~km away); and
\item local $g$ by a superconducting gravimeter (SG) at Apache Point. 
\end{itemize}
For daily estimates, the GPS positions at the P027 site typically
have uncertainties of about 1.5~mm horizontally, and 6~mm vertically.
Monthly averages---where LLR signatures are most relevant---have 0.3~mm
horizontal and 1.2~mm vertical uncertainties. Systematic errors prevent
these precisions from direct translation into accuracies for motions
relative to the Earth's center-of-mass, but the GPS data can still
usefully check models for site displacements. Fig.~\ref{fig:GPS}
shows data from the P027 site over four years, binned into periods
of 27.55 days (monthly periods are especially relevant to LLR science).
In the future, local GPS measurements may be used to constrain site
displacements in a simultaneous fit to LLR data.

Likewise, precision gravimetry can complement the vertically-challenged
GPS measurements by monitoring surface gravity variations. A superconducting
gravimeter mounted on the telescope pier of the Apache Point 3.5~m
telescope has the sensitivity on one-minute timescales to resolve
0.1~mm vertical displacements by virtue of the fact that surface
gravity diminishes by 3~nm~s$^{-2}$ for every millimeter of motion
away from the center of the Earth.

\begin{figure}[tbh]
\begin{center}\includegraphics[%
  width=6in,
  keepaspectratio]{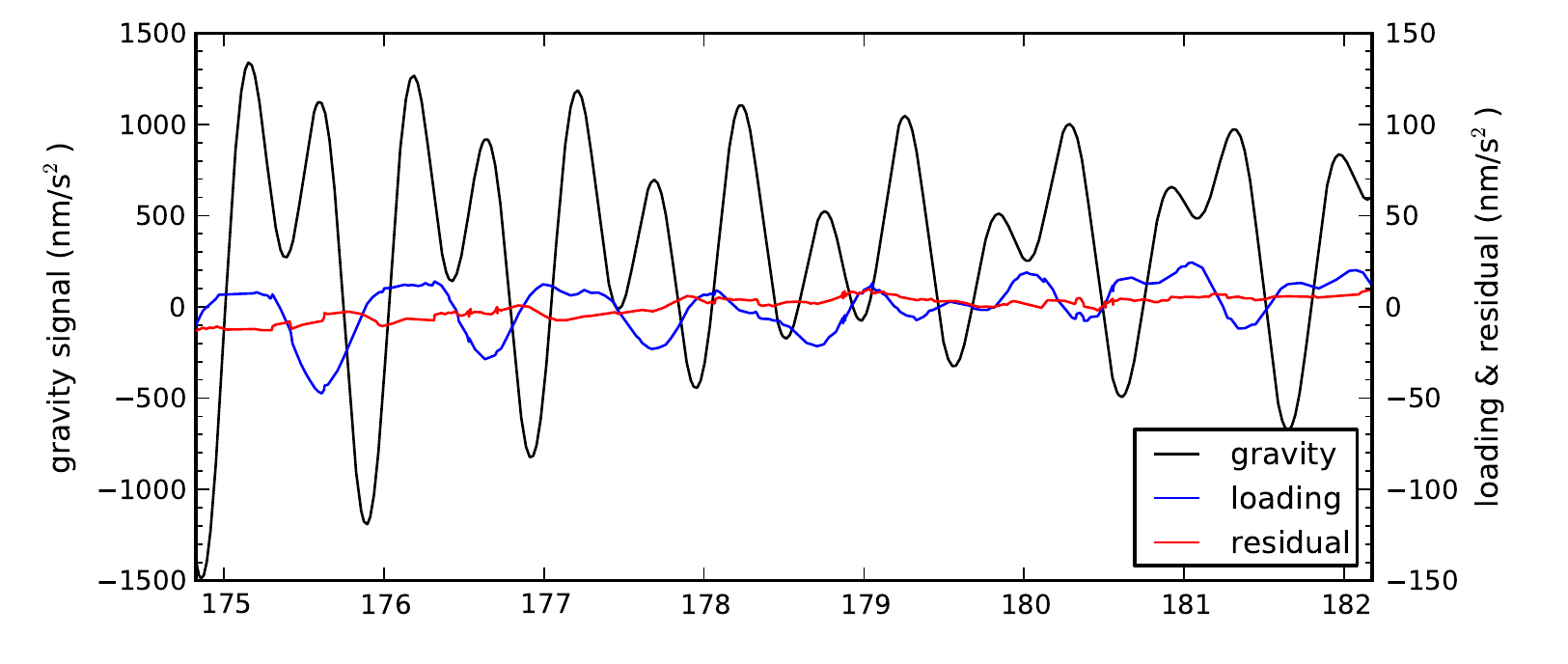}\end{center}

\caption{{\small One week of SG data at Apache Point. The harmonic (black)
line is not a model, but raw SG data filtered to one-minute samples.
The semi-periodic (blue) line---referenced to the right-hand scale
at 10$\times$ magnification---is the residual after subtracting a
tidal model and local atmospheric influence, but not ocean or hydrologic
loading signals. The flatter (red) curve is after removing ocean loads.
Steps/jumps visible in the red line are attributed to rotations of
the gravitationally asymmetric telescope dome, which can be subtracted
using a record of dome motion. A gravity deviation of 1~nm/s$^{2}$
corresponds to 0.18~mm of displacement for tides, and 0.3~mm for
ocean loading, meaning that the peak-to-peak motion represented here
amounts to more than 0.5~m. \label{fig:SG}}}
\end{figure}

The SG data measure local gravity variations with very low noise,
excellent calibration stability, and low instrument drift. When the
1~Hz sampled data are filtered to remove microseisms (typically 20~nm/s$^{2}$
peak-to-peak; 5--15~s periods), the short-term noise level is less
than $0.3$~nm/s$^{2}$ peak-to-peak, corresponding to 0.1~mm in
vertical displacement. Fig.~\ref{fig:SG} shows tides dominating
the SG signal, but removal of tides and local atmospheric influence
leaves a 60~nm/s$^{2}$ peak-to-peak signal dominated by ocean loading,
which when removed reveals small signals such as the gravitational
influence of the rotating telescope dome.

But a measurement of surface gravity only serves as a \emph{proxy}
to displacement. Direct gravitational attraction of loading sources
(atmosphere, ground water) complicate the picture, and the mass redistribution
accompanying tidal displacements change the potential---and therefore
the gradient of the potential, which is the measured quantity. For
tidal displacements, knowledge of the Love numbers, $h$ and $k$,
and the deforming potential, $W$, allows conversion from a measurement
of the variation in gravitational acceleration, $\Delta g$, to a
vertical displacement, $\Delta z$, via $\Delta g=-(1+h-\frac{3}{2}k)\frac{\partial W}{\partial r}$,
and $\Delta z=(1+k-h)W/g$. In principle, knowledge of the Love numbers
could come from the SG and/or the LLR fit. A better source may be
the recently much-improved elastic models for the Western U.S. resulting
from the U.S.A. Array seismometer project. But some uncertainty remains
in extrapolating from seismic frequencies to periods around a month. 

A modeling challenge of the future may be to apply the tidal model
to simultaneously fit the LLR data and gravimetry data. A demonstrated
ability to model the SG data---especially the long-period terms---will
provide an important estimate of the uncertainty from loading that
can be included in fits to the data. Part of the SG campaign involves
occasional comparison to a visiting absolute gravimeter in order to
calibrate long-term drift in the SG instrument. 

In a similar vein, while atmospheric propagation delay is well described
by measuring pressure, temperature, and humidity at the observing
site and applying a mapping function to the elevation of the observation
\cite{mendes-zenith,mendes-mapping}, horizontal pressure gradients
may foil the usual single-point pressure measurements. Using regional
pressure data (also useful for atmospheric loading corrections) has
been demonstrated to improve results \cite{atmos-gradient}. If greater
precision is required, incorporation of data from a kilometer-scale
barometric array may be employed to probe pressure gradients near
the observatory---which may be impacted by wind interacting with geographical
features, leading to dynamic pressure effects in the vicinity.

\subsection{Periodicity and Data Span\label{sub:periodicity}}

In this section, we look at temporal aspects of the LLR measurement and its
related science goals.  In the face of the large list of phenomenological
influences on the fundamental LLR measurement outlined in the preceding
sections, it is important to remember that most of the science goals
outlined in Section~\ref{sec:Science-from-LLR} rely on \emph{periodic}
range signatures. Clearly the equivalence principle signal is periodic,
displaying a $\cos D$ form at a period of 29.530589 days. This is likewise
true for gravitomagnetism and certain preferred frame effects. But even
secular effects ultimately derive from periodic observables. For instance,
a variation in $G$ alters the Keplerian relationship between period and
semi-major axis.  The resulting secular change in period results in a
quadratic phase evolution. Obviously LLR provides a nearly direct
measurement of the semi-major axis, but it is also directly sensitive to
the phase of the 21,000~km amplitude periodic variation due to orbital
eccentricity.  Likewise, sensitivity to precession of the orbit is provided
by monitoring the phase of the large-amplitude periodic behaviors in the
orbit.

The key point is that largely aperiodic phenomena like atmospheric
loading, variations in atmospheric propagation delay, or anomalous
meanderings in Earth orientation are unlikely to mimic science signals
at key lunar orbit periodicities. Obviously, greater sensitivity to
interesting science signals will derive from doing the best possible
job modeling confounding influences, effectively lowering the background
against which to seek small anomalous periodic signatures.

Even with this in mind, in order to take full scientific advantage of the
remarkable precision offered by LLR, the data must extend over a long
enough span to sample long-period terms in the lunar orbit so that
secular/aperiodic trends can be distinguished from periodic signatures and
optimum amplitudes of the various periodic signatures can be obtained. This
generally requires a data span of at least a half-period---and preferably
longer---of the longest important periodic term. The various monthly
periods: draconic\footnote{The draconic month describes the mean time it
takes for the Moon to cross the ecliptic plane in the ascending direction.}
(nodal passage: 27.212208 days); sidereal\footnote{The sidereal month is the
time it takes for the Moon to return to the same direction in inertial
space relative to the Earth center.} (inertial space period: 27.321661
days); and anomalistic\footnote{The anomalistic month refers to the lunar
mean anomaly, or phase/angle with respect to perigee.} (perigee-to-perigee:
27.554551 days) combine to give periods of 6.00 years, 8.85 years, and 18.6
years. While the complete LLR record now spans two of these longest
periods, not all of these data are of comparable precision. There is also a
75~yr wobble in the physical libration with a $\sim70$~m amplitude
\cite{rambaux}.

How might we expect formal uncertainties to scale with data span?
For the periodic signals associated with the equivalence principle,
gravitomagnetism, and preferred-frame effects, the uncertainty should
scale as $T^{-0.5}$ if the sampling is approximately uniform in time,
where $T$ is the data span. For secular drift of the lunar orbital
phase due to geodetic precession, deviations from the inverse square
law, or extra-dimension-motivated precession effects, sensitivity
scales as $T^{-1.5}$, where the additional power of $T$ comes from
a longer baseline. For $\dot{G}$, which effectively gauges secular
change of the orbital radius against that of the orbital period, the
uncertainty scales as $T^{-2.5}$, where two powers of $T$ come from
the aforementioned quadratic evolution of phase. These scalings only
apply to the extent that modeling capabilities are able to take full
advantage of the measurement precision.

\section{Future Advances\label{sec:Future-Advances}}

We discuss here potential future directions for LLR, together with
a qualitative assessment of resulting scientific gains.

Lunar laser ranging has for decades stayed at the forefront of tests
of gravity, probes of the lunar interior, and determination of Earth
coordinate systems. Recent improvements to the technique have stimulated
a push to improve modeling capabilities, which are expected to produce
further gains in the short term. Additionally, the LLR enterprise
has largely been confined to the northern hemisphere. Steady data
flow from a southern hemisphere station would allow better coverage
of low-declination observations and better constrain Earth orientation.

\subsection{Next-Generation Reflectors\label{sub:Next-Generation-Reflectors}}

Longer-term, improvements at the lunar end offer the biggest advantage---in
the form of either new reflectors, an active transponder, or both.
The current reflectors limit performance in a compounded way. Most
fundamentally, the finite extent of the reflector array spreads the
temporal width of the pulse by virtue of the fact that the array normal
tilts away from the line of sight by up to 10$^{\circ}$ due to lunar
libration (Figs.~\ref{fig:librations} and \ref{fig:libration-histogram}).
This spread can be as large as one nanosecond for the Apollo~15 array
at full-tilt, corresponding to a root-mean-square measurement uncertainty
approaching 50~mm (330~ps; see Fig.~\ref{fig:A15-RMS}). Statistical
centroiding of the signal to millimeter-level range precision requires
hundreds or thousands of photons. This is how APOLLO reaches the millimeter
domain, but such an approach is not feasible for other LLR stations.
Degraded reflector performance (Section~\ref{sub:Reflector-Degradation})
only exacerbates this problem. Thus the brute force approach to LLR
by gathering more photons becomes more challenging with time.

Meanwhile, the spread imposed by the tilted reflector array eliminates
incentives to improve ground-based laser pulse width or timing systems
in any incremental fashion, since these errors add in quadrature to
the dominant reflector spread. Improving APOLLO's 100~ps laser pulse
width and 20~ps timing system---even by a factor of two---would have
little discernible impact on the net timing precision, and so would
appear to be wasted effort. Installing a larger array on the Moon
also has no effect, as doubling the linear dimension doubles the temporal
spread, requiring four times the signal for statistical reduction
to the same level---which is exactly what a double-sized array delivers:
no precision gain.

Simply making a sparse array of corner cubes so that each one could
easily be resolved by $\sim100$~ps laser pulses would break the
logjam. Improvements in ground systems would then have immediate impact.
Halving the laser pulse width would consequently require four times fewer
photons for similar statistical precision. Most locations on the front
face of the Moon see the Earth permanently well away from local zenith,
so that a modest lateral separation on the ground ($>10$~cm) is
sufficient to separate the returns unambiguously. 

A few current efforts are underway to explore next-generation reflectors
for the lunar surface. Hollow corner cubes are being explored at the
Goddard Space Flight Center using an ultra-stable quartz bonding technique
\cite{preston}. Somewhat further along, work on 100~mm diameter
fused silica corner cubes is in the space-environment testing phase
to verify mitigation of thermal gradients in the presence of solar
illumination \cite{currie-lrrr}.

\subsection{Transponders\label{sub:Transponders}}

Installing active laser transponders on the lunar surface would have
perhaps an even greater impact on LLR science. Replacing the $1/r^{4}$
signal loss regime with a far more benign $1/r^{2}$ regime would allow the
extensive SLR network to engage in LLR on a routine basis.  This would have
tremendous impact in data volume, global distribution (fixing the southern
hemisphere deficit, for instance), tie-in to well-established geodetic
stations, and improvements in Earth surface/atmospheric models by using the
Moon as a reference object largely unaffected by non-gravitational
forces---unlike satellites. In this context, transponders work best in
asynchronous mode, rather than echoing detected incoming signals. This
permits the transponder to transmit a steady pulse train tied to a good
clock while recording times of incoming signals with respect to this clock.
The asynchronous mode has much greater noise immunity and thermal stability
than echo-based techniques.  Such transponders also pave the way for
interplanetary laser ranging.  The scientific benefits and hardware
requirements for one such system has been explored in the context of laser
ranging to Phobos \cite{phobos}.

\subsection{Impact on Science\label{sub:Future-Impact}}

On its face, improvement of LLR measurement precision as facilitated by new
reflectors, transponders, and concomitant ground station upgrades
has the potential to sharpen our constraints on (or find deviations in)
gravitational physics by a corresponding (equal) factor.  Time scales for
improvement vary for different science parameters, as discussed in
Section~\ref{sub:periodicity}, but in principle a factor-of-ten reduction
in LLR measurement uncertainty over a timescale of years to a decade
has the potential to deliver factor-of-ten improvements in LLR science.

Yet recent experience demonstrates that improving LLR measurement precision is
not by itself sufficient to realize scientific gains.  The model must also
keep pace.  Sections~\ref{sub:Charted-Improvements} and
\ref{sub:Uncharted-Improvements} provide a glimpse into the host of
phenomena
one must consider in plotting a course from centimeter-level to millimeter-level
LLR accuracy.  Progressing into the sub-millimeter regime will undoubtedly
invoke a similarly-sized---if not larger---list of concerns that may or may
not be tractable.  Even so, the aforementioned reflector/transponder
upgrades would permit a greater global distribution of ever-improving
LLR-capable stations, in addition to relieving the current burden on
statistical reduction of the dominant tilt-induced timing uncertainty.  Since many of the challenges confronting LLR today
relate to Earth phenomenology, widespread global participation may be the
best way to characterize these influences and reduce their impact on LLR
science goals.  Improved reflectors and/or transponders may provide the
most robust route for future improvements in LLR science. 

\section{Conclusion}

Since its inception, LLR has established itself as a mainstay of precision
measurement relating to gravitation, physics of the Earth-Moon system,
and coordinate systems. The basic measurement is general enough to
have broad reach across many dimensions of physics. Within the gravity
sector, LLR provides the very best probes of the equivalence principle,
the time-rate-of-change of the gravitational constant, gravitomagnetism,
geodetic precession, the inverse square law, preferred frame effects,
and is also well positioned to test new ideas in physics. For the
Earth-Moon system, dissipative processes in the lunar interior expose
a liquid core, LLR measurements contribute to knowledge of Earth orientation
and coordinate systems, and tidal dissipation on Earth is seen via
the egress of the lunar orbit. 

Despite order-of-magnitude degradation of reflector signal strength,
LLR data are now gathered at unprecedented rates and approaching one-millimeter
range precision as a consequence. Rediscovery of the long-lost Lunokhod~1
reflector brings the total available reflectors on the Moon up to
five. At this time, the lunar reflectors are the limiting source of
temporal uncertainty in the ranging error budget, so that new reflectors
on the lunar surface would offer a dramatic improvement in range precision
capability. Additionally, new reflectors or even transponders on the
lunar surface could open up LLR to dozens of satellite laser ranging
stations around the world, vastly improving data volume, global distribution,
and interest in the science.

Alongside the challenge of acquiring accurate lunar range measurements,
the sophisticated model that accounts for every relevant influence
must see concomitant improvements. Currently in the process of adapting
to millimeter-quality data after decades of centimeter-quality measurements,
newly improved limits on science from LLR may be around the corner.

\ack

Some of the text in this article was adapted from proposals, to which
Eric Adelberger, Christopher Stubbs, John Chandler, Duncan Agnew,
and David Crossley contributed. Some text has also been adapted from
the author's direct contribution to the review article by M\" uller
et al. \cite{jog-rev}, which also includes U. Schreiber, P. J. Shelus,
J.--M. Torre, J. G. Williams, D. H. Boggs, S. Boquillon, and G. Francou
as co-authors. Some text is adapted from the APOLLO instrument paper
\cite{apollo}, with co-authors E. G. Adelberger, J. B. R. Battat,
L. N. Carey, C. D. Hoyle, P. LeBlanc, E. L. Michelsen, K. Nordtvedt,
A. E. Orin, J. D. Strasburg, C. W. Stubbs, H. E. Swanson, and E. Williams.
Many of the results presented here reflect efforts by the same list
of individuals associated with the APOLLO project. Nathan Johnson
offered helpful comments. The author acknowledges funding support
for lunar ranging and associated efforts by the National Science Foundation
(Grant PHY-1068879), by NASA (NNX12AE96G and NNX10AU99G) and by the
NASA Lunar Science Institute as part of the LUNAR consortium (NNA09DB30A).

\appendix

\section*{Acronyms Used in Text}

{\small
\begin{tabular}{ll}
APD	& Avalanche Photodiode \\
APOLLO	& Apache Point Observatory Lunar Laser-ranging Operation \\
AU 	& Astronomical Unit ($1.496\times 10^{11}$~m) \\
CfA	& Harvard-Smithsonian Center for Astrophysics \\
CSR4	& Center for Space Research ocean loading model 4 \\
DGP	& Dvali, Gabadadze, Porrati \\
DORIS	& Doppler Orbitography and Radiopositioning Integrated by Satellite \\
EIH	& Einstein-Infeld-Hoffmann \\
EP	& Equivalence Principle \\
FWHM	& Full-Width and Half-Maximum \\
GOT00	& Global Ocean Tide ocean loading model \\
GPS	& Global Positioning System \\
GR	& General Relativity \\
GRACE	& Gravity Recovery and Climate Experiment (Earth) \\
GRAIL	& Gravity Recovery and Interior Laboratory (Moon) \\
ICRF	& International Celestial Reference System \\
IERS	& International Earth Rotation and Reference System Service \\
IMCCE	& Institut de mecanique celeste et de calcul des ephemerides \\
JPL	& Jet Propulsion Laboratory \\
LAGEOS	& Laser Geodynamics Satellites \\
LLR	& Lunar Laser Ranging \\
LOLA	& Lunar Orbiter Laser Altimeter \\
LRO	& Lunar Reconnaissance Orbiter \\
LUNAR	& Lunar University Network for Astrophysical Research \\
MLRS	& McDonald Laser Ranging System \\
NASA	& National Aeronautics and Space Administration \\
OCA	& Observatoire de la C\^ote d'Azur \\
PEP	& Planetary Ephemeris Program \\
PPN	& Parameterized Post Newtonian \\
RMS	& Root-Mean-Square \\
SBR	& Signal to Background Ratio \\
SEP	& Strong Equivalence Principle \\
SG	& Superconducting Gravimeter \\
SLR	& Satellite Laser Ranging \\
SME	& Standard Model Extension \\
SSB	& Solar System Barycenter \\
TDC	& Time to Digital Converter \\
TPXO	& TOPEX/Poseiden-based ocean loading model \\
UT1	& Universal Time offset 1 \\
UTC	& Coordinated Universal Time \\
VLBI	& Very Long Baseline Interferometry \\
WEP	& Weak Equivalence Principle \\
YAG	& Yttrium aluminum Garnet \\
\end{tabular}
}

\end{document}